\documentclass[lettersize,journal]{IEEEtran}
\usepackage{amsmath,amsfonts} 
\usepackage{amsthm,amssymb}
\usepackage{algorithm}
\usepackage{algorithmic}
\usepackage{setspace}
\usepackage{array}
\usepackage[table]{xcolor}
\usepackage{makecell} 
\usepackage{multirow}
\usepackage{booktabs}
\usepackage{graphicx}  
\usepackage{tabularx}  
\usepackage{epstopdf}
\usepackage{subcaption} 
\usepackage{textcomp}
\usepackage{wasysym}
\usepackage{mathrsfs}
\usepackage{stfloats}
\usepackage{float}
\usepackage{breakcites} 
\usepackage{cleveref}
\usepackage{url} 
\usepackage{ulem}
\usepackage{verbatim}
\usepackage{pifont}  
\usepackage{lipsum} 
\usepackage[most]{tcolorbox}
\usepackage{colortbl}
\usepackage{epstopdf}
\begin{document}
\title{SVAFD: A Secure and Verifiable Co-Aggregation Protocol for Federated Distillation}

\author{Tian Wen, Sheng Sun, Yuwei Wang, ~\IEEEmembership{Member,~IEEE}, Peiyan Chen, Zhiyuan Wu, Min Liu, ~\IEEEmembership{Senior Member,~IEEE}, Bo Gao, ~\IEEEmembership{Member,~IEEE} }

\maketitle
\begin{abstract}
Secure Aggregation (SA) is an indispensable component of Federated Learning (FL) that concentrates on privacy preservation while allowing for robust aggregation. However, most SA designs rely heavily on the unrealistic assumption of homogeneous model architectures. \textbf{\underline{F}}ederated \textbf{\underline{D}}istillation (FD), which aggregates locally computed logits instead of model parameters, introduces a promising alternative for cooperative training in heterogeneous model settings. Nevertheless, we recognize two major challenges in implementing SA for FD. (i) Prior SA designs encourage a dominant server, who is solely responsible for collecting, aggregating and distributing. Such central  authority facilitates server to forge aggregation proofs or collude to bypass the claimed security guarantees; (ii) Existing SA, tailored for FL models, overlook the intrinsic properties of logits, making them unsuitable for FD. 

To address these challenges, we propose SVAFD, the first SA protocol that is specifically designed for FD to enable both privacy protection, logits integrity and verifiability. At a high level, SVAFD incorporates two innovations: (i) a multilateral co-aggregation method tha redefines the responsibilities of both clients and server. Clients autonomously evaluate and aggregate logits shares locally with a lightweight coding scheme, while the powerful server handles ciphertext decoding and performs the computationally intensive task of generating verification proofs; (ii) a quality-aware knowledge filtration method that facilitates biased logits exclusion against poisoning attacks. Moreover, SVAFD is resilient to stragglers and colluding clients, making it well-suited for dynamic networks in real-world applications. We have implemented the SVAFD prototype over four emerging FD architectures and evaluated it against ten types of poisoning and inference attacks. Results demonstrate that SVAFD improves model accuracy and guarantees secure and robust aggregation, making it a significant step forward in secure and verifiable aggregation for heterogeneous FL systems.
\end{abstract}


\section{Introduction}
\IEEEPARstart{R}{ecent} years have witnessed growing interest in Federated Learning (FL), a collaborative Machine Learning (ML) paradigm that enables model training without compromising the confidentiality of native datasets \cite{letaiefEdgeArtificialIntelligence2022,Wu2023AgglomerativeFL}. Leveraging parallel computation and enhanced privacy, FL has been increasingly applied many domains, supporting efficient AI deployment\cite{lim2020federated}. However, its broader adoption is hindered by: 1) High communication overhead: The periodic exchange of model parameters entails frequent communication, with the overhead increasing proportionally to model size; and 2) Homogeneous model constraint: Typical FL enforces a uniform model architecture across clients, which is incompatible with diverse computational capabilities or application requirements of participants\cite{wuExploringDistributedKnowledge2023}.

\begin{figure}[!t]
  \centering
  \includegraphics[width=1\linewidth, height=0.5\textheight, keepaspectratio]{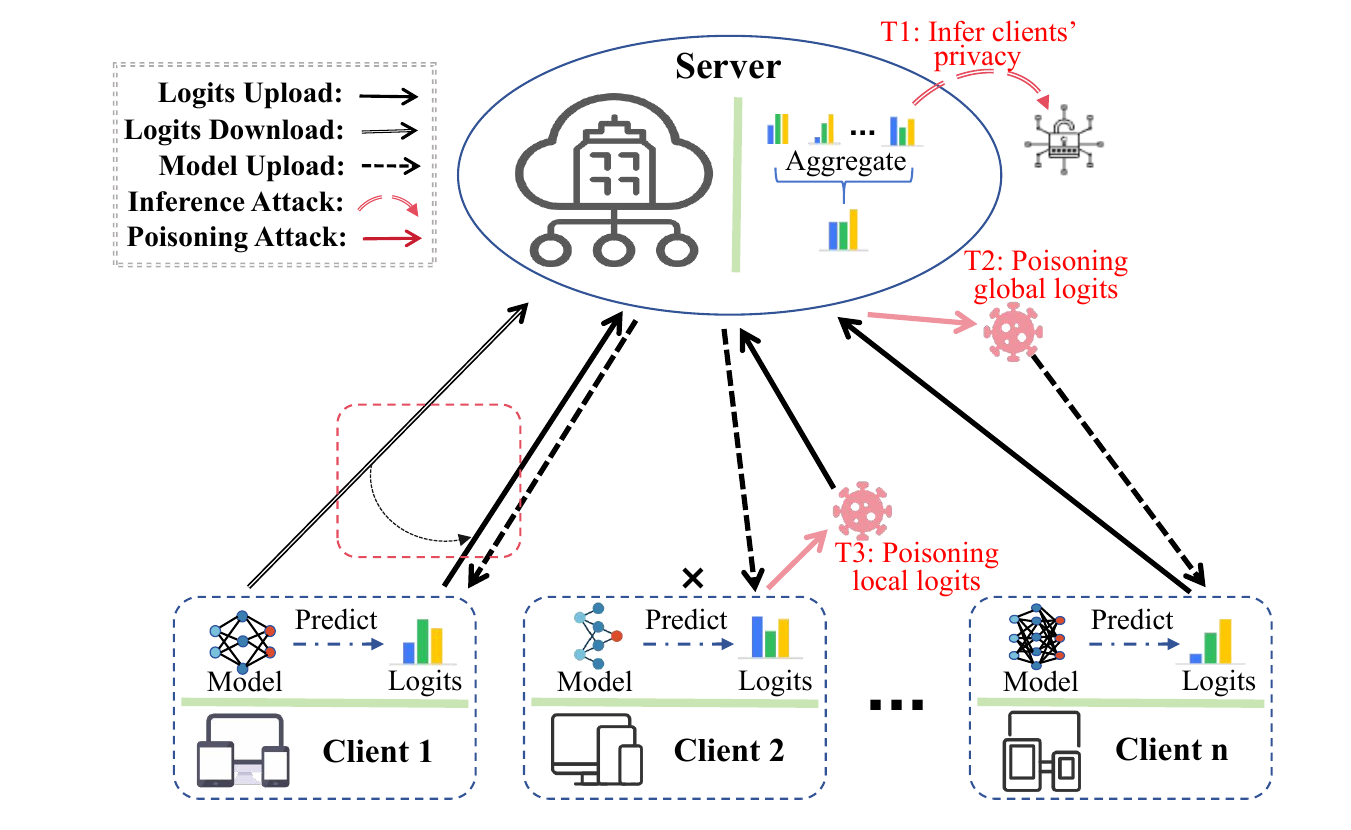}
  \caption{Federated Distillation Architecture and Security Threats. Note: The red box highlights that in FD, clients upload logits instead of models as in FL; the symbol "T" combined with a number denotes specific security threats, where threats "T1" and "T2" occur on the server side, while threat "T3" manifests on the client side.}
  \label{fig1}
\end{figure}

Federated Distillation (FD), an improved paradigm of Federated Learning (FL), overcomes the aforementioned limitations by exchanging model outputs (i.e., logits) instead of full model parameters or updates\cite{wu2023survey}.  As illustrated in Fig.\ref{fig1}, each client generates and uploads local logits to a central server, which aggregates them into teacher knowledge (called "global logits") to guide subsequent local training. Compared to conventional FL, FD significantly reduces communication overhead, as it depends only on the dimension of model outputs rather than the size of the entire model. In addition, FD naturally supports heterogeneous model training, allowing clients to adopt personalized architectures and thereby offering greater flexibility and efficiency\cite{9964434}.

The growing adoption of FD systems necessitates a thorough examination of their robustness and security. Three types of security threats show in Fig.\ref{fig1} underscore that both the server and the clients can be corrupted by adversaries, leading to mainly two categories of malicious attacks\footnote{These attacks persist in FL scenarios, however, this paper focuses exclusively on the FD scenarios.}: i) \textit{Inference attacks}, targeting the violation of participants' data privacy to capture distribution of data features\cite{shaoSelectiveKnowledgeSharing2023} or even achieve pixel-level reconstruction of training data \cite{takahashiBreachingFedMDImage2023}; 
ii): \textit{Poisoning attacks}, aiming at undermining the integrity of global logits, which are performed through generating misleading model or fabricated local logits\cite{ratheeELSASecureAggregation2023}. While the Secure Aggregation(SA) countermeasures has been extensively studied in conventional FL\cite{lycklamaRoFLRobustnessSecure2023,PracticalSecureAggregationa}, achieving robust and efficient SA in FD settings, particularly under heterogeneous model architectures, remains an open and underexplored issues. This paper specifically focuses on advancing secure aggregation under such heterogeneous FD scenarios.

\textbf{Challenges:} we recognize two major challenges for realizing SA within FD. Firstly, previous SA protocols are vulnerable to a dominant server\cite{ratheeELSASecureAggregation2023}. As shown in Table.\ref{table1}, the SA protocols, which simultaneously achieve both privacy protection and system integrity, either rely on a trusted third party\cite{ratheeELSASecureAggregation2023,yuEVFLEfficientVerifiable2024} or resort to a semi-honest server\cite{lycklamaRoFLRobustnessSecure2023,liMartFLEnablingUtilityDriven2024}. These protocols encourage a dominant server to act as the aggregation subject, controlling all processes, including knowledge filtering, knowledge aggregation, and necessary authentication for knowledge distribution. However, \cite{gao2023vcd} indicates that the server, by colluding with just one client, can easily forge credentials or decrypt sensitive client information, thus undermining the secure guarantees claimed by the existing SA. Secondly, prior SA protocols are susceptible to clients' poisoning attacks\cite{Tang2024LogitsPA,Tang2024PeakControlledLP,10646975}, and the works on mitigating this issue can be roughly categorized into malicious detection approaches or byzantine-robust estimators introducing non-linear aggregation rules with audits to filter biased updates\cite{lycklamaRoFLRobustnessSecure2023,ratheeELSASecureAggregation2023}. These SA protocols primarily rely on observing the characteristics of model weights/updates, with some theories suggesting that parameter updates from benign clients exhibit continuity and consistency across successive training rounds, and the malicious behaviors can be detected by monitoring changes in model weight patterns, update directions, or norm constraints\cite{zhaoFederatedLearningNonIID2022}. However, there is no research demonstrating similar characteristics of logits in FD across training rounds. The knowledge filtration method that considers the intrinsic properties of logits have been studied to a lesser extent.

\begin{table}[t]
\centering
\caption{The Overall Evaluation of Reviewed Solutions in terms of privacy, integrity, verifiable and trusted setting with $\checkmark$ and $\times$ presenting the ability. SH refers to semi-honest, and M denotes the malicious server.}
\label{table1}
\renewcommand{\arraystretch}{1.3}
\begin{tabularx}{0.475\textwidth}{c|c|c|c|c}
\hline
\makebox[0.1\textwidth][c]{\textbf{SA Protocols}}
 & \makebox[0.07\textwidth][c]{\textbf{Privacy}} & \makebox[0.07\textwidth][c]{\textbf{Integrity }} & \makebox[0.07\textwidth][c]{\textbf{Verifiable}} & \makebox[0.06\textwidth][c]{\textbf{Server}}  
 
 \\ \hline
AHEFL\cite{8241854} & $\checkmark$ &  $\times$ & $\times$ &SH  \\ \hline
MKFL\cite{jiangSecureNeuralNetwork2021} &$\checkmark$ &  $\times$ & $\times$ &   SH \\ \hline
VPFL\cite{shenVerifiablePrivacyPreservingFederated2023} & $\checkmark$&  $\times$& $\checkmark$ &    SH \\ \hline
PSAML\cite{PracticalSecureAggregationa} & $\checkmark$ &  $\times$&  $\times$ &    M  \\ \hline
PolySA\cite{bellSecureSingleServerAggregation2020} & $\checkmark$ &  $\times$ &  $\times$&   M \\ \hline
FLShiled\cite{kabirFLShieldValidationBased2024} & $\checkmark$ & $\checkmark$ &  $\times$ &  SH \\ \hline
RoFL\cite{lycklamaRoFLRobustnessSecure2023} & $\checkmark$ & $\checkmark$ &  $\checkmark$ &   SH\\ \hline
ELSA\cite{ratheeELSASecureAggregation2023} & $\checkmark$ & $\checkmark$ &  $\times$ &  SH \\ \hline
martFL\cite{liMartFLEnablingUtilityDriven2024} & $\checkmark$ & $\checkmark$ & $\checkmark$ &   SH \\ \hline
\textbf{Ours} & $\checkmark$ &$\checkmark$ &  $\checkmark$ &  M \\ \hline 
\end{tabularx}
\end{table}

\begin{figure*}[!t]
  \centering
  \includegraphics[width=0.8\linewidth]{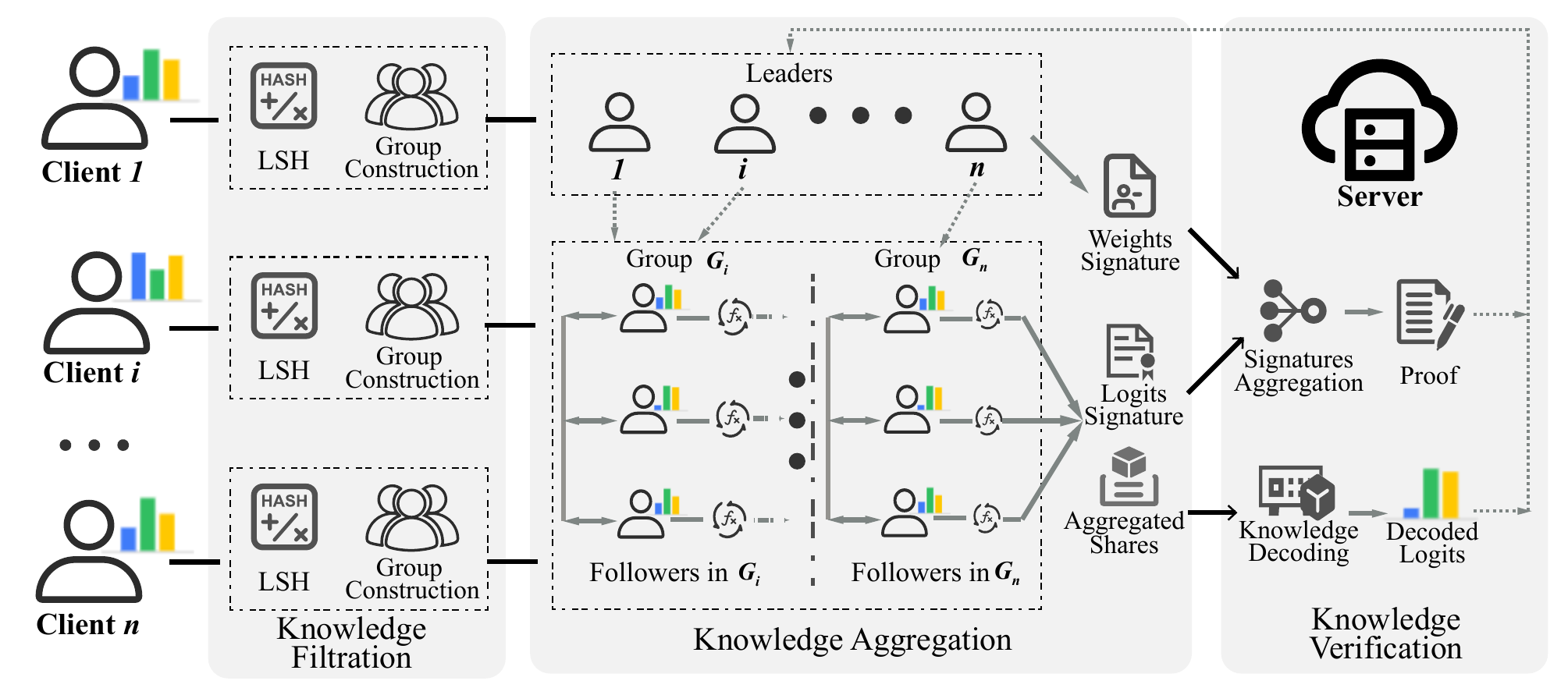}
  \caption{Overview of SVAFD.}
  \label{fig2}
\end{figure*}

\textbf{Proposed defense:} To fill the research gap, we begin by examining the properties of logits and outline the central issue of this paper: \textbf{how to construct a verifiable SA protocol that simultaneously ensures privacy protection and knowledge integrity for FD?} We propose SVAFD, a novel verifiable secure co-aggregation protocol specifically designed for FD as illustrated in Fig.\ref{fig2}. SVAFD decouples the SA process into three distinct stages: knowledge filtration, knowledge aggregation, and knowledge verification. First, clients autonomously filter knowledge based on the similarity of logits distribution. Then, using lightweight Lagrange Coding Computation (LCC) encoding, clients aggregate the encoded knowledge shares locally. Ultimately, the server decodes the aggregated results to recover teacher knowledge and applies bilinear pairings to aggregate knowledge signatures, providing proof for the execution of the aggregation process. 

SVAFD shifts the aggregation process from unilateral server dominance to multilateral co-aggregation, and each party assumes specific responsibilities. While clients independently select and aggregate local logits with appropriate weights, the server focuses on decoding the aggregated results and generating signatures, which is a process that involves significant computational resources. Moreover, SVAFD incorporates features such as straggler tolerance and privacy protection against colluding inference, making it well-suited for deployment in resource-constrained edge networks. 

\textbf{Contribution:}
The main contribution of this paper is the design, implementation, and evaluation of SVAFD, the first SA protocol tailored for FD scenario. SVAFD ensures privacy protection, knowledge integrity, and verifiability in the presence of a malicious server. We have implemented the SVAFD prototype consisting of approximately 1,250 lines of code, which will be provided after the Review Stage. The prototype successfully adapts to four emerging FD architectures, results demonstrate that it significantly improves both accuracy and security. It effectively resists eight types of poisoning attacks and one type of inference attacks, achieving high accuracy and low attack success rates. Furthermore, we provide a detailed analysis of the system-level overhead, demonstrating that SVAFD is efficient and practical for real-world deployment.

\section{PRELIMINARY}

\subsection{Overview of FD}
We investigate the Federated Distillation (FD) architecture, where distributed devices (called clients) collaborate to train a $D$-class classification model coordinated by the server. Without losing generality, we consider the FD system with $N:=\{1,2,\cdots,n\}$ clients. Each client $n \in N$ holds a private dataset $\mathcal{S}_n:= \bigcup_{i=1}^{s_n} {(X^n_i, y^n_i)}$ which includes $s_n = |\mathcal{S}_n|$ data samples, with $x^n_i$ and $y^n_i$ respectively indicating the feature vector and the correspinding label of the $i$-th training sample.  Different from the homogeneous model settings of conventional FL system \cite{kabirFLShieldValidationBased2024,lycklamaRoFLRobustnessSecure2023,ratheeELSASecureAggregation2023,liMartFLEnablingUtilityDriven2024}, we expect that each client $n$ maintains a personalized model $\mathcal{M}^n := (\Phi^n, \ell^n)$, which differs in model parameters $\Phi^n$ and architectures. $\ell^n(\cdot)$ is the non-linear mapping determined by $\mathcal{M}^n$. The optimization objective of the FD system is to maximize the average User Accuracy (UA) \cite{wuFedCacheKnowledgeCachedriven2023a,wuFedICTFederatedMultitask2023,mills2021multi} across all clients, that is to achieve generally satisfactory performance on each client. 

In contrast to sharing model parameters or updates, FD employs the knowledge guided mechanism that is communication-efficient and heterogeneity-friendly to enable collaborative training. The server aggregates local knowledge from multiple clients to construct global knowledge, which is subsequently disseminated back to the clients to enhance local training.  Throughout this paper, local knowledge is defined as equivalent to the logits $\vartheta_n$, representing the raw, unnormalized predictive confidence generated by the local model $M^n$ of client $n$  on the local dataset $\mathcal{S}_n$.  

Depending on the granularity of logits exchanged in each round of Client-Server (C/S) interactions, existing FD architectures can be classified into two categories\cite{wuExploringDistributedKnowledge2023,wuFedCacheKnowledgeCachedriven2023a}: the Class-grained Logits Interaction-based FD Ar-
chitecture (CLIA) \cite{jeongCommunicationEfficientDeviceMachine2018} with logits $\vartheta_c\in \mathbb{R}^{D \times D}$ capturing class-level predictive capabilities, and the Sample-grained Logits Interaction-based FD Architecture (SLIA) \cite{wuFedCacheKnowledgeCachedriven2023a, liFedMDHeterogenousFederated2019a,itaharaDistillationBasedSemiSupervisedFederated2023}, where logits $\vartheta_c\in \mathbb{R}^{O \times D}$ represent sample-level predictions, with $O$ denoting the number of distilled samples drawn from the public dataset \cite{wuExploringDistributedKnowledge2023}. SLIA enables finer-grained logits exchange. A detailed discussion of these two logits granularity architectures is provided in Appendix \ref{app:A}.

The process of FD training can be divided into two stages, outlined as follows:
\begin{itemize}

\item \textbf{Local Train.} Each client $n \in N$ optimizes its local model parameters $\Phi^n$ a composite loss function, which combines the cross-entropy loss $L_{CE}(\cdot)$ for the local dataset $S_n$ with the distillation loss $L_{KD}(\cdot)$ derived from the global logits, i.e.:
\begin{equation}\label{eq1}
\begin{aligned}
\min _{\Phi^n} \underset{\left(x_i^n, y_i^n\right) \sim \mathcal{S}^n}{E}\left[J_{C E}+\lambda \cdot J_{K D}\right]
\end{aligned}
\end{equation}
which is subject to:
\begin{equation}\label{eq2}
\begin{aligned}
\left\{\begin{array}{l}
\left.J_{C E}=L_{C E}\left(\ell^n\left(\Phi^n ; X_i^n\right), y_i^n\right)\right) \\
J_{K D}=L_{K D}\left(\ell^n\left(\Phi^n ; X_i^n\right), Y\left(X_i^n\right)\right)
\end{array}\right.
\end{aligned}
\end{equation}
where $\lambda$ represents the distillation weight factor. $Y(\cdot)$ refers to the global logits received from the central server in the latest round. This optimization produces the updated local model that generates the new local knowledge, represented as logits $\vartheta_n$.

\item \textbf{Global Aggregation.}  
After collecting all the clients' logits $\{\vartheta_n|n \in N\}$, the server performs an aggregation operation to obtain the global logits as:
\begin{equation}\label{eq4}
\begin{aligned}
Y=\hbar(\ \{\vartheta_n| n\in N \} )
\end{aligned}
\end{equation}
where $\hbar(\cdot)$ is the aggregation operation managed by the server. The two aforementioned processes iterate continuously until a predefined number of rounds or a target accuracy is achieved.
\end{itemize}

\subsection{Lagrange Coding Computation}
The Lagrange Coding Computation (LCC) \cite{yuLagrangeCodedComputing2019} represents a sophisticated encoding technique crafted to facilitate secure computations within distributed environments, striking a balance between computing efficiency and data privacy.\cite{asheralievaAuctionLearningBasedLagrange2023,soleymaniAnalogLagrangeCoded2021}. The key motivation of LCC is to encode the pending data using Lagrange polynomials that enable computational redundancy and plaintext inaccessibility across the distributed works , thereby preserving data privacy and enabling straggler tolerance \cite{NEURIPS2022_448fc91f}.  

As illustrated in \ref{fig3}(a), the Data Provider (DP) holds a dataset $S_0$, on which a target function $f(S_0)$ is to be computed, where $f(\cdot)$ represents either linear transformations or complex polynomial operations. To enable privacy-preserving and distributed computation, the DP first encodes $S_0$ into $N$ shares using LCC, and distributes these encoded shares to a set of $N$ Service Processors (SPs), i.e., the clients. Each DP performs local computation $f(\cdot)$ on its assigned share and returns the corresponding partial result to the DP. Finally, the DP decodes the collected results to reconstruct the global output $f(S_0)$. The fundamental $(D,T)$-achievable encoding property of LCC is outlined in \textbf{Theorem 1}, with a formal proof provided in Section IV of \cite{yuLagrangeCodedComputing2019}. 

\textbf{Theorem 1}. \textit{Given $N$ Data Processors (DPs) and the degree $\text{deg}(f)$ of the  target function $f(\cdot)$, a D-resilient and T-private Lagrange Coding Computation (LCC) is achievable, as long as}
\begin{equation}\label{eqnew}
\begin{aligned}
D + deg(f)(K+T-1)+1\leq N
\end{aligned}
\end{equation}
where $K$ denotes the number of slices that the pending dataset $S_0$ splited. $D$-Resilient indicates that even with up to $D$ clients dropping out, the server can decode the global output $f(S_0)$. $T$-Private ensures that, even if up to $T$ clients collude, they cannot infer any meaningful information about $S_0$. 

\subsection{Threat Model and Basic Requirements}
In SVAFD, we assume that all clients agree to the knowledge aggregation process facilitated by a central server, and consent to share the essential data of knowledge and signature authentication to complete the aggregation. SVAFD contrasts with that in \cite{PracticalSecureAggregationa}, where an end-to-end information relay is managed by the central server via a KPI\cite{PracticalSecureAggregationa,bellSecureSingleServerAggregation2020}. Instead, it adopts the communication protocol similar to \cite{kalraDecentralizedFederatedLearning2023,soltaniDFLStarDecentralizedFederated2024}, where client devices can establish direct communication channels with other clients. Since mobile devices can only sporadically access power and network connections, the set of participants in each update round is unpredictable and can join or leave at any time.

Based on the above assumptions, we consider the server to be malicious in terms of reliability. The server may arbitrarily deviate from the protocol by returning incorrect proof or poisonous knowledge to honest clients. This assumption captures the adversarial behavior in multi-party collaborative computing, where the server's goal is to deceive users into accepting incorrect results. Meanwhile, the clients are semi-malicious in terms of reliability, which are required to honestly provide CAL (described in \S \ref{3.3}) for intimacy computation and sign the split knowledge slices. Additionally, there is no assumption of honesty regarding the clients, who may submit arbitrary biased data at any stage of interaction to launch poisoning or inference attacks that disrupt the FD training process.

In terms of privacy, both of the server and clients are malicious, and may collude to achieve the best attack capability in capturing victims' privacy. We assume that the cryptographic primitives used for knowledge encoding in SVAFD are secure, and the bilinear mapping used in our verifiable knowledge aggregation is reasonable. We believe that the above assumptions are stricter and more realistic than the existing trusted settings.

\textbf{Basic requirements for SA:} This paper delves into the SA problem in the FD scenario, with particular emphasis on three paramount security attributes:
\begin{enumerate}
    \item \textbf{Privacy protection} Even in the face of a malicious server and up to $T$ malicious clients colluding in inference attacks, the logits privacy of honest clients is guaranteed.
    
    \item \textbf{Knowledge integrity:} Fully resilient to poisoning attacks from a malicious server, while increasing tolerance to poisoning attacks from malicious clients.
    
    \item \textbf{Verifiability:} Provide clients with verification of knowledge aggregation execution. No malicious clients or server, even when colluding, can generate the proof to deceive victims.
\end{enumerate}

\section{Challenges and Key Insights of SVAFD}
\subsection{Challenges of Secure Aggreagation (SA)}

We analyze the reasons behind security threats from the server and client sides, respectively. 

\begin{figure*}[!t]
  \centering
  \includegraphics[width=1\linewidth]{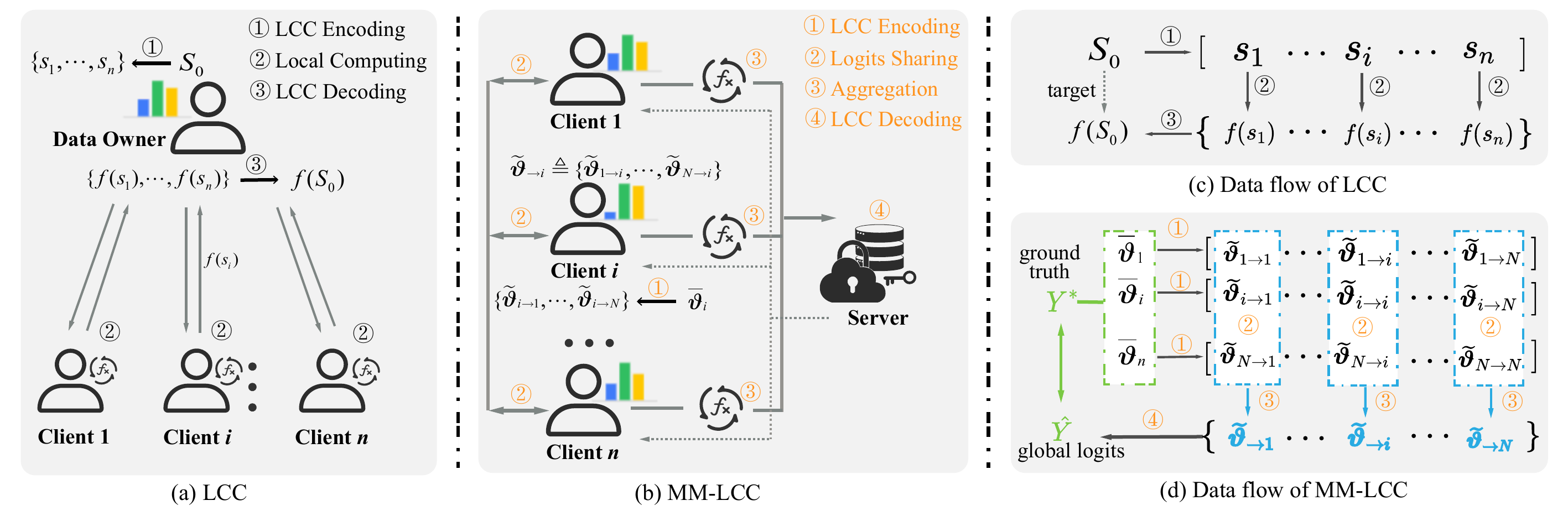}
  \caption{Workflow of  LCC and MM-LCC.}
  \label{fig3}
\end{figure*}

\textbf{\uline{Challenge \ding{172}}:} \textbf{Existing FL/FD architectures rely on a dominant server as the aggregation subject, which may facilitate server-side poisoning and inference attacks. }

The current FL/FD architectures designate the server as the active coordinator of aggregation, with full access to the model/knowledge uploaded by clients and control over the distribution of the global model/knowledge. Although this design is efficient in coordinating clients, it creates opportunities for potential security threats. Regarding inference attacks, the academic community has reached a basic consensus that the server should not have access to the plaintext information about any client' knowledge or model.  A series of studies based on encryption and perturbation have emerged, achieving breakthroughs in efficiency and performance. However, most of these approaches resort to the week security assumption of an honest-but-curious server and lack scalability to address malicious server poisoning attacks. Solutions to counter poisoning attacks need to provide clients with aggregation verification, often employing techniques such as homomorphic hashing, commitments, polynomial verification, or zero-knowledge proofs. These methods either introduce unacceptable verification overhead on client-side or fail to prevent collusion between server and clients. In other words,  as long as a single client colludes with the server, the verification conditions can be forged.

The high privileges of the dominant server result in inherent flaws in most existing SA architectures. A revolutionary approach adopts a fully distributed end-to-end aggregation framework. However, without centralized server coordination, achieving global synchronization (e.g., consistency checks or progress coordination) becomes extremely challenging, and multi-party security authentication as well as model performance improvement cannot be guaranteed. Neither server-dominated nor fully distributed client-driven SA protocols offer an optimal solution.

This empirical evidence underscores a critical insight:

\begin{tcolorbox}[
    colback=blue!5!white,    
    colframe=blue!40!black, 
    boxrule=0.4pt,           
    arc=2mm,              
    left=1mm,                
    right=1mm,          
    top=1mm,      
    bottom=1mm     
    ]
 
\textbf{\textcolor{blue!60!black!}{Plaintext invisibility and auditable aggregation  should be ensured. }}  \textit{To address security threats originating from the server, it is necessary to simultaneously guarantee clients’ privacy and enable collusion-resistant verification with acceptable computational overhead. }
\end{tcolorbox}

In light of this, our defense strategy, SVAFD, introduces a verifiable co-aggregation method based on VMLCC. Clients first encode and share local logits using lightweight LCC. Subsequently, each client performs local aggregation on the received knowledge slices and submits the aggregation results, along with the signature information, to the server for global decoding and signature aggregation. SVAFD shifts the aggregation process from unilateral server dominance to multilateral co-aggregation. Clients are responsible for encoding their logits, ensuring end-side privacy, while the server handles global decoding and signature aggregation to guarantee the verifiability of the aggregation process. Experimental results show that SVAFD does not introduce excessive computational overhead on the client side but instead offloads most computational tasks to servers with stronger resource capabilities. More details are deferred to \S \ref{3.2}.

\textbf{\uline{Challenge \ding{173}}:} \textbf{Current malicious client detection methods are primarily designed for FL and are not readily applicable to FD. }

In FL, model weight updates $\Delta w_t$  from clients typically exhibit the following properties: \textit{Continuity}, $\lVert\Delta w_t - \Delta w_{t-1}\rVert \leq \epsilon$, meaning that the difference between updates in adjacent rounds is relatively small; \textit{Consistency}, the update $\Delta w_t^{i}$ from a benign client tends to align in direction with the global aggregated update $\Delta w_t$.

Due to the fundamental differences in the nature of model updates and knowledge transfer in FD, we underscore a critical insight:

\begin{tcolorbox}[
    colback=blue!5!white,    
    colframe=blue!40!black, 
    boxrule=0.4pt,           
    arc=2mm,              
    left=1mm,                
    right=1mm,          
    top=1mm,      
    bottom=1mm     
    ]
 
\textbf{\textcolor{blue!60!black!}{Malicious detection method should be specifically tailored to FD on the integrity of logits.}}  \textit{To effectively address security threats originating from clients in FD settings, a deeper analysis on the statistical properties and filtration method of logits is necessary.  }
\end{tcolorbox}

Our proposed methodology, SVAFD, introduces a novel defensive mechanism that strategically employs clients within high-qualy knowledge, specifically based on the statistical properties of logits.
More details are deferred to section \S \ref{3.3}.

\subsection{MM-LCC Design for Co-aggregation}
\label{3.2}
Multi-to-Multi LCC (MM-LCC), including multi Data Providers with multi Service Providers, is an extension of the traditional One-to-Multi LCC. Viewing MM-LCC as a primitive, we could construct a strawman co-aggregation workflow as shown in Fig.\ref{fig3}(b). In this diagram, each client acts as a DP, with the public objective of obtaining the aggregated global logits of all clients, i.e: 
\begin{equation}\label{eq5}
\begin{aligned}
\hbar(\mathcal{V})=\mathcal{W}f(\mathcal{V})
\end{aligned}
\end{equation}
where $\mathcal{V} = [\vartheta_1, \dots, \vartheta_c]$ and $\mathcal{W}=[w_1,\cdots,w_c],c \in [N]$ denote the local logits and aggregation weights set, respectively. At the same time, each client also serves as the SP, required to provide slice aggregation services for DPs. 

At a high level, MM-LCC shifts the aggregation process from \textbf{unilateral server dominance}, where the server fully controls the aggregation and distribution of global knowledge, to \textbf{multilateral co-aggregation}, where all clients are no longer passive participants but instead locally aggregate knowledge slices, which are then submitted to the server for global logits decoding.  The dataflow of MM-LCC is illustrated in Fig.\ref{fig3}(d). Each client $c$, 1) utilizes LCC to encode the local logits $\vartheta_c$ into multiple shares $\{\widetilde{\vartheta}_{c\rightarrow i}|i \in [N]\}$ for outsourced computing; 2) shares the logits splits with each other, where $\widetilde{\vartheta}_{c\rightarrow i}$ denotes the logits slice sent from client $c$ to client $i$; 3) performs the aggregation operation $\hbar(\widetilde{\vartheta}_{\rightarrow c}) $ on the received logits shares $\widetilde{\vartheta}_{\rightarrow c}=[\widetilde{\vartheta}_{1\rightarrow c},\cdots,\widetilde{\vartheta}_{i\rightarrow c}]$, shown in the blue box in Fig.\ref{fig3}(d). Finally, the server, possessing abundant computational resources, 4) decodes the aggregated shares set $\{ \hbar(\widetilde{\vartheta}_{\rightarrow c})| c \in N\}$ to obtain the global teacher knowledge $\hat{Y}$.

We define the following properties that the MM-LCC satisfied to support secure co-aggregation:

\textbf{Definition 1: (D,T,$\hbar$)-achievable.} \textit{MM-LCC enables secure co-aggregation for its property of  D-Resilient, T-Privacy and $\hbar$-operational, as long as $D + deg(\hbar)(K+T-1)+1\leq N$.}

Given $N$ Clients and the degree $\text{deg}(\hbar)$ of the aggregation function $\hbar(\cdot)$, 
D-Resilient donotes that the server can decode the global logits $\hat{Y}$ even with up to $D$ clients dropping out. This is because MM-LCC inherently adopts the redundant encoding scheme used by LCC. T-Privacy eliminates the leakage of clients' privacy, demonstrating that no meaningful information about local logits $\vartheta_n, n \in N$ can be inferred by the curious server or at most $T$ colluding clients. Ultimately, $\hbar$-operational highlights any aggregation functions with the degree of deg($\hbar$) supported. It indicates that MM-LCC not only accommodates linear operations in FD with $deg(\hbar)=1$, but is also extendable to scenarios requiring intricate polynomial mappings for high-dimensional feature representations, such as general tensor algebra, gradient computation\cite{NEURIPS2022_448fc91f} and bilinear computation\cite{yuLagrangeCodedComputing2019}.

\begin{figure}[!t]
  \centering
  \subfloat[Updating direction with different logits.]{\includegraphics[width=0.45\linewidth]{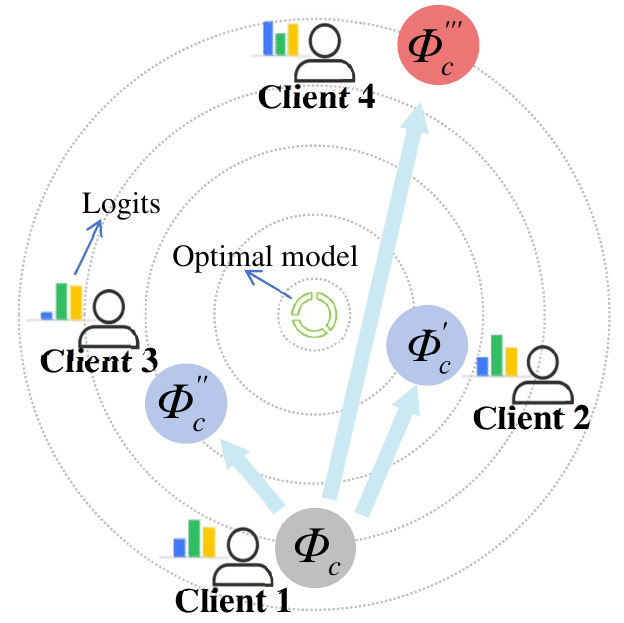}}
  \hspace{0.5cm} 
  \subfloat[Similaries between logits and data distribution.]{\includegraphics[width=0.45\linewidth]{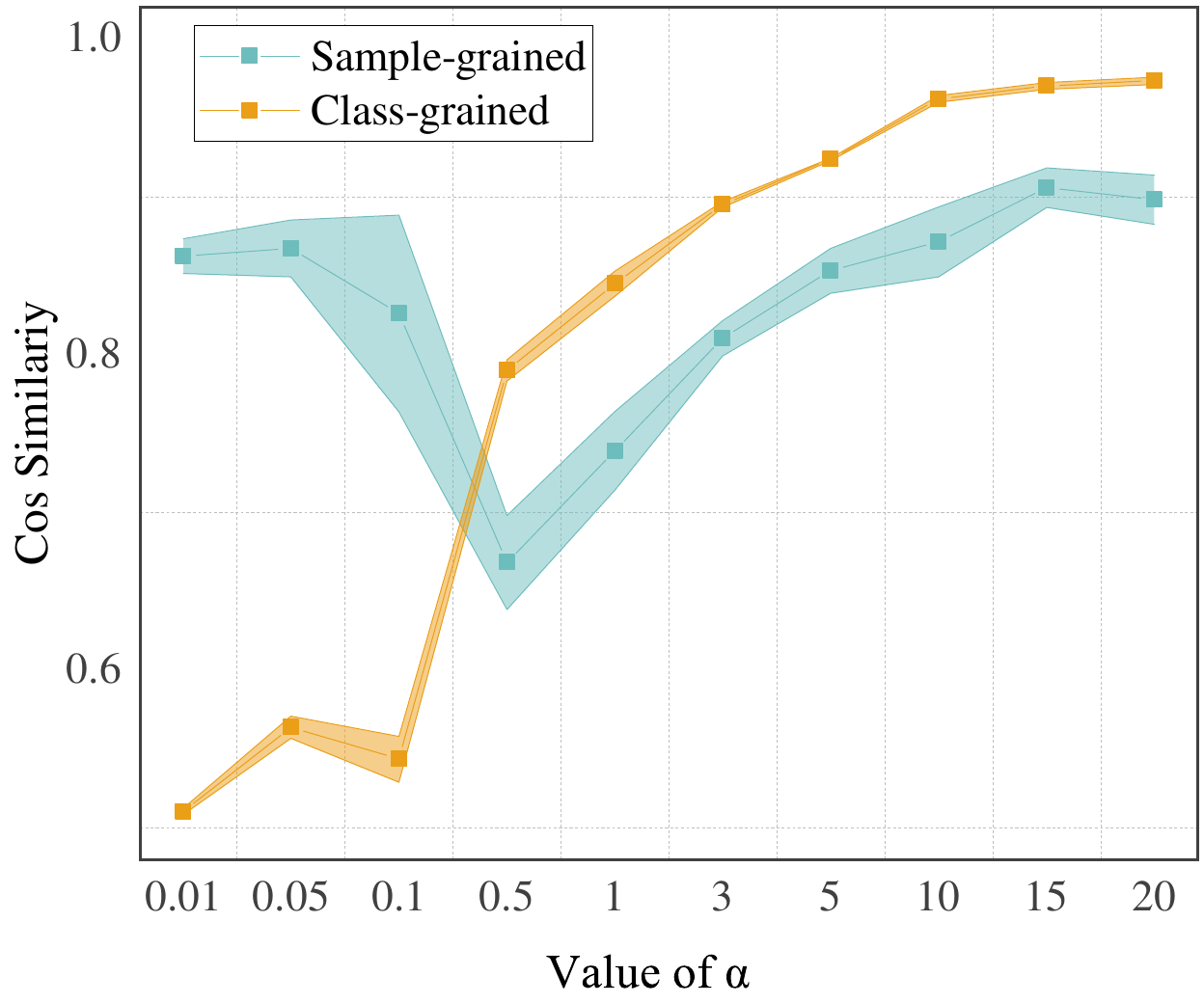}}
  \caption{Observation of CAL}
  \label{fig4}
\end{figure}

\textbf{Definition 2: Homomorphic Share Aggregation.} \textit{MM-LCC is utilized to enable co-aggregation for its property of  Homomorphic Share Aggregation. Specifically, any operation performed by the client on the knowledge shares, after decoding, must correspond to the direct operation on the original knowledge data: }

\begin{equation}\label{eq6}
\begin{aligned}
\hbar(\mathcal{V})=\operatorname{Dec(\hbar(\operatorname{Enc(\mathcal{V})}))}
\end{aligned}
\end{equation}
where the $Enc(\cdot)$ and $Dec(\cdot)$ represent the LCC encoding and decoing, respectively.

Definition 2 provides a theoretical basis for end-side share aggregation and server-side global decoding. It indicates that the global knowledge  $\hat{Y}$ obtained through decoding is equivalent to the ground truth value $Y^*=\mathcal{W}f(\mathcal{V})$, illustrated in the green color of Fig.\ref{fig3}(d).

\subsection{CAL for Quality-aware Knowledge Filtration}
\label{3.3}
The model on each client tends to learn personalized representations based on an independently sampled space, which favors higher-frequency data samples, thereby enhancing local fitting accuracy \cite{wuExploringDistributedKnowledge2023}. 
Given the heterogeneity of the models and the data distribution, the logits from each client often exhibit significant differences. Fig.\ref{fig4}(a) illustrates the model update direction after client 1 respectively aggregates knowledge with the other three clients. Clients 1, 2, and 3 exhibit similar logits, allowing effective guidance towards optimization along the gradient descent direction. However, due to the knowledge inconsistencies, the aggregation of client 1 and 4 results in biased logits, which in turn affects the optimization rate and convergence direction.

Aggregating logits from heterogeneous clients without any discrimination results in biased knowledge representations\cite{shaoSelectiveKnowledgeSharing2023}. A straightforward solution is to select a subset of clients with closely local knowledge. Fig.\ref{fig4}(b) illustrates the average cosine similarity between the mean logits uploaded by a set of clients and their respective local data distributions, where $\alpha$ representes the heterogeneity of Dilliclet distribution. It suggests that logits are highly correlated with the data distributions of the clients. Furthermore, in accordance with Hinton’s analysis\cite{hintonDistillingKnowledgeNeural2015}, logits provide rich information regarding the probability distributions of each class and the relationships between different classes.  These insights help us with logits filtration. Therefore, we introduce a new quality aware metric, class average logits (CAL), which is computed class-wise to evaluate the knowledge distribution of the client. Our experiments reveal that employing CAL in the knowledge filtration process causes benign aggregated teacher knowledge to be more consistent with each other. We formally define the CAL metric as follows:
\begin{equation}\label{eq7}
\begin{aligned}
\mathcal{C}^c=[\frac{1}{\underset{\left(x_i^c, y_i^c\right) \in \mathcal{S}^c}{\sum}\delta_{[y_i^c=y_d]}} 
\underset{\left(X_i^c, y_i^c\right) \in \mathcal{S}^c}{\sum} \delta_{[ y_i^c=y_d]} \ell^c\left(x_i^c\right) \\ | \ y_d\in [1,c] \ ] 
\end{aligned}
\end{equation}
where $\delta$ is a Boolean variable that equals 1 when $y_d = y_i^c$, and 0 otherwise. $\ell^c\left(x_i^c\right)$ refers to the logits related to
sample $x_i^c$. Then the CAL $\mathcal{C}^c,c\in [N]$ is shared among clients using the Locality-Sensitive Hashing (LSH)\cite{10.5555/645925.671516} function $\mathcal{{L}}:\mathbb{R}^{D\times D}\rightarrow \mathbb{R}^{D \times P}$, which ensures that similar CAL are mapped to the same hash bucket with high probability. $P$ is the hyperparameter representing the size of hashed value.

We utilize the cosine similarity to filter the intimacy for each pair of clients, as follows:
\begin{equation}\label{eq8}
\begin{aligned}
\mathcal{A}(c, z) = \frac{\mathcal{L}(\mathcal{C}^c)\cdot \mathcal{L}(\mathcal{C}^z)}{\|\mathcal{L}(\mathcal{C}^c)\| \cdot \|\mathcal{L}(\mathcal{C}^z)\|}
\end{aligned}
\end{equation}
where $\|\cdot\|$ denotes the norm. By simplifying $\mathcal{A}(c, z)$ as $a_{c,z}$, each client calculates its affinity list vector $A_{c}=[a_{c,1}, a_{c,2}, \cdots, a_{c,n}]$ and selects the most intimate clients set $G_c$ for knowledge aggregation.

\textbf{Note 1:} 
Each client $c\in [N]$ assumes two roles: one as the leader, requesting for aggregation of all $G_c$ members to generate the teacher knowledge, and the other as a follower, contributing its local knowledge to the respective group leaders.

\section{IMPLEMENTATION OF SVAFD}

\subsection{Framework Overview}
SVAFD proposes a secure knowledge co-aggregation method that balances privacy protection, stability, and verifiability. After each round of local training by the clients, SVAFD starts to implement with three stages: knowledge filtration, knowledge aggregation, and knowledge verification.

In Knowledge Filtration, clients exchange Local CAL through the LSH, and establish their personalized intimacy groups to select benign followers, respectively. Specifically, each client $c\in [N]$ first generate the $\mathcal{C}^c$ across all data samples and then maps it to hashed value $\mathcal{L}({\mathcal{C}^c})$ for similarity computation. Then client $c \in [N]$, as the leader, can select the top $R$ most intimate clients to establishes the group $G_c$ for teacher knowledge aggregation. Meanwhile, the communication topology of all groups is constructed into a global network graph, which is dynamically updated at each training round.

\begin{figure}[!t]
  \centering
  \includegraphics[width=1\linewidth]{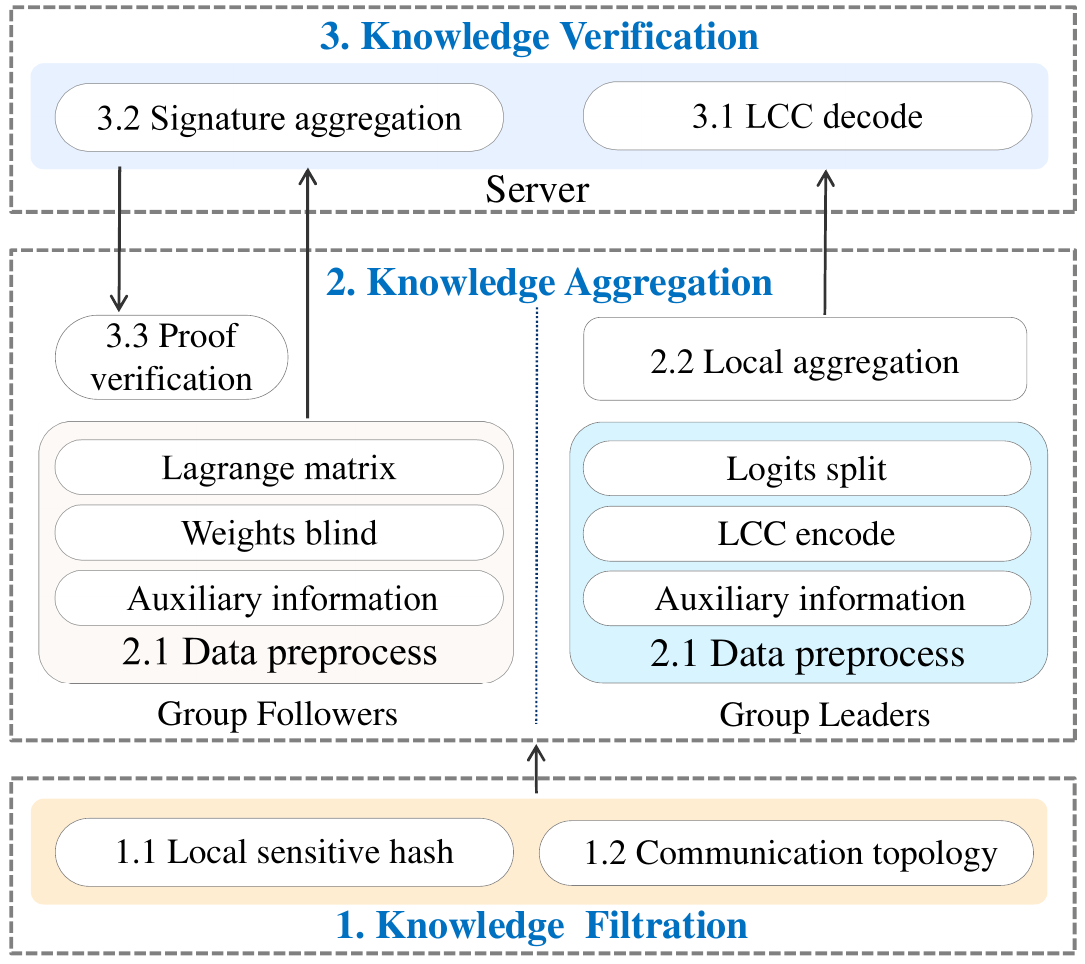}
  \caption{Workflow of SVAFD.}
  \label{fig5}
\end{figure}

In Knowledge Aggregation, group leaders initialize the aggregation weights and Lagrange coefficients for their followers. The followers, in turn, encode the local logits, exchange the encoded logits, and aggregate the received shares locally. For each group $G_c$, the leader client $c$ distributes the LCC coefficient matrix $L_c$ used for encoding and the masked weights $\widetilde{\mathcal{W}}_c$ to its group member $z \in G_c$. Followers first split their local logits to obtain $\overline \vartheta$ and joint noise matrices. Then, they encode $\overline \vartheta$ by LCC, deliver the corresponding logits share with each other, and aggregate all the received shares respectively. Meanwhile, both the leader and followers generate auxiliary signature information $(\pi_{\mathcal{V}}, \pi_{\mathcal{W}})$, which will later upload to the server. Additionally, all followers send their private signature keys $\Upsilon_z$ to the leader. In Knowledge Verification, the server is responsible for decoding the received aggregated shares $\hbar(\widetilde \vartheta)$ and generating the verification proof by bilinear pairings for all group leaders. After receiving the decoded teacher knowledge $\widetilde{Y}_{c}$ and $\pi_c$, the leader decodes and authenticates the teacher knowledge for next round's training. The workflow of SVAFD is shown in Fig.\ref{fig5}.

\subsection{Knowledge Filtration}

\subsubsection{Local Sensitive Hash}
Since the CALs are calculated according to Eq.\ref{eq7}, the $\mathcal{C}^c$ values of client $c\in [N]$ is firstly projected by the local sensitive hashing $\mathcal{L}(\cdot)$ to obtain $\mathcal{L}(\mathcal{C}^c)$. Then, clients distribute their local private hashed values with each other, and calculate the intimacy list $A_{c}=[a_{c,1},a_{c,2},\cdots,a_{c,{n}}]$ based on the received $\{\mathcal{L}(\mathcal{C}^c) \mid c \in [N] \}$ according to Eq.\ref{eq8}. The top $R$ most similar candidates are selected to form the group as follows:
\begin{equation}\label{eq9}
    \begin{aligned}
    G_c=\{z \mid z\in \hat{Top}(A_c,R),z\neq c\}
    \end{aligned}
\end{equation}
where $\hat{Top}(A_c,R)$ selects R followers with the largest values from $A_c$.

\textbf{Note 2:} While the leader $c \in [N]$ selects its followers to form the group $G_c$, it can also be chosen as a follower to join other groups. 

\subsubsection{Communication Topology} Let $\mathcal{G}_c = \{ G_x \mid c \in G_x, x \in [N] \} \cup \{G_c\}$ be the set of groups that client $c \in [N]$ participates in. Given the set $\mathscr{G} = \{\mathcal{G}_c \mid c \in [N]\}$, we construct the communication topology $\mathit{G} = (N, \mathcal{E})$, where $N$ represents the set of client nodes, while $\mathcal{E} =\{(c, z) \mid 1 \leq c \leq [N], z \in \bigcup_{G_x \in \mathcal{G}_k} G_k\}$ representing the set of edges.

\textbf{Note 3:} The definition in $\mathcal{E}$ indicates that once a client joins a group (whether following or leading), it will establish a communication channel with other group members. A limited number of accessible groups effectively prevents resource-constrained nodes from establishing too many communication connections, that leading to communication congestion or single points of failure. 

\subsection{Knowledge Aggregation}

\subsubsection{Data Preprocess}
\textbf{1) logits split:} We consider two forms of logits splitting methods.

Case 1: \textit{Class-grained Split}. The logits mode of client $z \in G_c$ is only related to the class numbers $D$, namely $\vartheta_z \in \mathbb{R}^{D \times D}$. 

With the additive splitting method, we split $\vartheta_z$ into $K$ parts to obtain $\overline{\vartheta}_z \triangleq [\overline{\vartheta}_z^{(1)}, \overline{\vartheta}_z^{(2)}, \dots, \overline{\vartheta}_z^{(K)}] \in \mathbb{R}^{K \times D \times D}$, and define the $(g,l)$-th element of $\vartheta_z$ and $\overline{\vartheta}_z^{(k)}$ as $v_{g,l}$ and $ v_{g,l}^{(k)}$, satisfying $ v_{g,l} = \sum_{k=1}^{K} v_{g,l}^{(k)}$, where $g,l\in [D]$. This split method also applicable to sample-grain FD, but for efficiency, the following method is proposed to handle the sample-grain FD\cite{liuPrivacyPreservingVerifiableOutsourcing2023}.

Case 2: \textit{Sample-grained split}. The logits mode is proportional to the related to the number of sample numbers $O$ of the distilled dataset, namely $\vartheta_z \in \mathbb{R}^{O\times D}$.

We deals with the logits by block split method. Each client $\vartheta_z$ is divided into $K$ shares $\overline{\vartheta}_z^{(k)} \in \mathbb{R}^{\frac{O}{K} \times D} $, with each block containing $ \frac{O}{K}$ samples. Without loss of generality, we assume that $O$ is an integer multiple of the batch size parameter, which is divisible by $K$ allowing for better and adaptation to batch processing strategy of GPUs. Then, local knowledge slice matrix  is represented $\overline{\vartheta}_z \in \mathbb{R}^{K \times \frac{O}{K} \times D} $. For ease of interaction in later steps, we unify the mode description of client logits with $\overline{\vartheta}_z \in \mathbb{R}^{K \times \Omega \times D}$ to represent their mode.

To ensure information-theoretic privacy, a uniform random matrix $\overline{H}_z  \triangleq [\overline{h}_z^{(1)}, \overline{h}_z^{(2)}, \dots, \overline{h}_z^{(T)}]$
is appended to blind the splited $\overline{\vartheta}_z$, where $\overline{h}_z^{(t)} \in \mathbb{C}^{\Omega \times D}$ is independently sampled from a zero-mean circularly symmetric complex gaussian distribution, with truncated standard deviation $[-\theta\frac{\sigma_z}{\sqrt{T}}, \theta\frac{\sigma_z}{\sqrt{T}}]$. The parameter $T$ measures privacy protection capability, representing the maximum number of colluding attackers allowed, $\sigma_z$ denotes the matrix variance, and $\theta$ is the truncation coefficient. The matrix concatenation operation is denoted by $\oplus$, then the blind operation $B(\cdot)$is defined as follows:

\begin{equation}\label{eq10}
\begin{aligned}
B(\overline{\vartheta}_z) &\triangleq [\overline{\vartheta}_{z}^{(1)}, \dots, \overline{\vartheta}_{z}^{(K)}] \oplus [\overline{h}^{(1)}_z, \dots, \overline{h}^{(T)}_z ]\\
&\triangleq [\overline{\vartheta_{i}}^{(1)}, \overline{\vartheta_{z}}^{(2)}, \dots, \overline{\vartheta_{z}}^{(K)}, \overline{h}^{(1)}_z, \dots, \overline{h}^{(T)}_z ] 
\end{aligned}
\end{equation}

\textbf{2) Lagrange matrix and weights blind:}
While followers in group $G_c$ performing logits splitting, the group leader $c$ will send them the Lagrange encoding matrix and masked weights for subsequent LCC encoding and aggregation operations.

As the Analog LCC\cite{soleymaniAnalogLagrangeCoded2021} encoding is employed, the leader computes the Lagrange coefficient by:
\begin{equation}\label{eq11}
    \begin{aligned}
        l_z(\alpha) = \prod_{l \in [K+T] \setminus \{ z \}} \frac{\alpha - \beta_l}{\beta_z - \beta_l}
    \end{aligned}
\end{equation}
for all $z \in [K+T]$. Additionally, the parameters $\alpha$ and $\beta$ are picked to be equally spaced on the circle of radius centered around 0 in the complex plane by $\alpha_i= \gamma^{i-1}, i \in [G_c]$ and $\beta_j =\omega^{j-1},j \in [K+T]$, where $\gamma = e^{-\frac{2\pi \iota}{R}}$ and $\omega = e^{-\frac{2\pi \iota}{K+T}}$ are the $N$-th and $(K+T)$-th roots of unity, respectively, with $\iota^2 = -1$. 
Based on the above elements, the Lagrange coefficient matrix for LCC encoding is generated as:

\begin{equation}\label{eq12}
    L_c \overset{\text{def}}{=} \begin{bmatrix}
l_1(\alpha_{1}) & \cdots & l_{K+T}(\alpha_{1}) \\
l_1(\alpha_{2}) & \cdots & l_{K+T}(\alpha_{2}) \\
\vdots & \ddots & \vdots \\
l_1(\alpha_{x}) & \cdots & l_{K+T}(\alpha_{x})
\end{bmatrix}^T_{R\times (K+T)}
\end{equation}

SVAFD supports each group leader customizing the aggregation weights $\mathcal{W}_c = [w_1,w_2 \dots, w_z], z \in G_c$, and chooses a random matrix $\textbf{R}_c \in \mathbb{R}^{\Omega \times D}$ as the factor to blind the weight matrix. We perform the product operation between the blinding factor and each element of $\mathcal{W}_c$ as
$
\widetilde{\mathcal{W}}_c= 
    [w_1\cdot \mathbf{R}_c ,w_2 \cdot \mathbf{R}_c ,\cdots,w_z \cdot \mathbf{R}_c]
\in \mathbb{R}^{R \times (\Omega \times D)}
$. And the $(L_c,\widetilde{\mathcal{W}}_c)$ is eventually distributed to each groud followers in $G_c$.

\textbf{3) LCC Encode:} Each followers $z \in G_c$ performs LCC encoding on the polynomial $u_z: \mathbb{C} \rightarrow \mathbb{C}^{\Omega \times D}$, interpolating at $\alpha_x, x \in G_c$ as:

\begin{equation}
\begin{aligned}\label{lcc}
u_z(\alpha_x)
&=\sum_{k=1}^K  \overline{\vartheta}^{(k)}_{z}l_k(\alpha_x) + \sum_{t=K+1}^{K+T} \overline{h}_{z}^{(t-K)} l_t(\alpha_x) \\
&=\widetilde{\vartheta}_{z \rightarrow x}
\end{aligned}
\end{equation}

where $\widetilde{\vartheta}_{z \rightarrow x}$ denotes the logits share from follower $z$ to $x$. Thus, the encoded logits shares sent by client $z$ are, $
\widetilde{\vartheta}_{z \rightarrow }=[\widetilde \vartheta_{z\rightarrow 1},\widetilde \vartheta_{z\rightarrow 2}, \cdots,\widetilde \vartheta_{z \rightarrow x}], z,x\in G_c
$

\textbf{4) Auxiliary Information}

The leader and followers in group $G_c$ need to compute auxiliary proofs $(\pi{'}, \pi^c)$, respectively. The $\pi{'}$ represents the auxiliary proof of the encoded logits slices by followers, while the latter $\pi^c$ is the auxiliary proof that leader generates for the aggregation weights $\mathcal{W}_c$. 

To strike a balance between privacy protection and efficiency, the follower $z$ calculates the sum of elements for each of the K slices in the splitted logits $\overline{\vartheta}_z = [\overline{\vartheta}_z^{(1)}, \overline{\vartheta}_z^{(2)}, \dots, \overline{\vartheta}_z^{(K)}]$ to get the matrix digest $[V_{z}^{(1)},V_{z}^{(2)},\cdots,V_{z}^{(K)}]$, where $V_{z}^{(k)} = \sum_{g=1}^{\Omega} \sum_{l=1}^{D} v_{g,l}^{(k)}$. Then, the auxiliary signatures of local logits can be computed as

\begin{equation}
\begin{aligned}\label{eq14}
\pi^{(k)'}_z=g^{V_z^{(k)}+\Upsilon_z},z \in G_c,k\in [K]
\end{aligned}
\end{equation}
where  $\Upsilon_z$ denotes the private key of the follower $z$, which helps the leader detect dishonest behavior by malicious actors during the SA. Then the client $c$ attatins the auxiliary proof $\pi_z'=\{\pi_z^{(1)'},\pi_z^{(2)'},\cdots, \allowbreak, \pi_z^{(K)'}\},z \in G_c$. In the mean time, the leader also signs the weights $[w_1,w_2,\cdots,w_z]$ by $\pi^{c}_z=g^{w_z}, z \in G_c$. Finally, the leader and followers will generate and then upload the auxiliary proof $\pi_{\mathcal{V}}^c = (\pi_1', \pi_2', \dots, \pi_z')$ and $\pi_{\mathcal{W}}^c = (\pi^c_1, \pi^c_2, \dots, \pi^c_z)$ to the server in Knowledge Verification stage.

\textbf{Note 4:} Due to the fact that most signature computation approaches are designed based on large integers,
we inplement the conversion of $ V_z^{(k)}$ and  $w_z$ through an approximate function as shown in Appendix \ref{app:C}. 

\subsubsection{Local Aggregation}

In this stage, followers in $G_c$ distribute the encoded logits sharings with each other. Ultimately, each follower $z \in G_c$ will receive a set of logits shares from all the $R$ clients as $\widetilde{\vartheta}_{\rightarrow z}=[\widetilde \vartheta_{1\rightarrow z},\widetilde \vartheta_{2\rightarrow z},\cdots,\widetilde \vartheta_{x\rightarrow z}]\in \mathbb{C}^{R \times \Omega \times D}, x\in G_c$

According to the Eq. \ref{eq6}, the knowledge aggregation operation $\hbar(\cdot)$ firstly applies a polynomial function $f(x)$ to each of the received shares, then performs the linear weighted sum to integrate the teacher knowledge shares. Specifically, for $z \in G_c$, the aggregation $\hbar: \mathbb{C}^{R \times \Omega \times D} \rightarrow \mathbb{C}^{\Omega \times D}$ denotes as

\begin{equation}
\begin{aligned}\label{eq15}
\hbar(\widetilde \vartheta_{\rightarrow z})=\widetilde W_c f(\widetilde{\vartheta}_{\rightarrow z})
\end{aligned}
\end{equation}

The set of all aggregated results from all followers in group $G_c$ can be represented as:
$\{ \hbar(\widetilde \vartheta_{\rightarrow z})\mid z \in  G_c\}$, which is then delivered to the server for decoding.

\subsection{Knowledge Verification}

\subsubsection{LCC Decode}
In this section, the server decodes the teacher knowledge by collecting sufficient numbers of aggregation results from each group. For group $G_c, c \in [N]$,  each of the the aggregated sharings $\hbar(\widetilde{\vartheta}_{\rightarrow z})$ corresponds to an evaluation of $ \hbar(u(\alpha_z))$, $z \in G_c$, and the $u(\cdot)$ is:

\begin{equation}\label{lccs}
\begin{aligned}
u(\alpha_z) 
&=\sum_{k=1}^K \overline{\mathcal{V}}^{(k)}l_k(\alpha_z) + \sum_{t=K+1}^{K+T} \overline{\mathcal{H}}^{(t-K)} l_t(\alpha_z) \\
\end{aligned}
\end{equation}
where
$$
\begin{aligned}
\overline{\mathcal{V}}^{(k)} &= [\overline{\vartheta}_{1}^{(k)}; \overline{\vartheta}_{2}^{(k)}; \cdots; \overline{\vartheta}_{z}^{(k)}] \in \mathbb{R}^{R \times \Omega \times D} \\
\overline{\mathcal{H}}^{(t-K)} &= [\overline{h}_{1}^{(t-K)}; \overline{h}_{2}^{(t-K)}; \cdots; \overline{h}_{z}^{(t-K)}] \in \mathbb{C}^{R \times \Omega \times D}
\end{aligned}
$$

The decoding process consists of two steps. Firstly, the server needs to recover the coefficients of the polynomial $ \hbar(u(\alpha_z))$. Since the degree of this composite polynomial is $deg(\hbar)(K + T - 1)$, the server requires at least $deg(f(x))(K + T - 1) + 1$ aggregated shares (evaluation points) to interpolate and construct the polynomial, where $deg(\hbar)=deg(f(x))$. Secondly, the server evaluates the polynomial $\hbar(u(z))$ at point $z = \beta_k, k\in [K]$ to obtain $\overline{Y}_c^{(k)}=\hbar(\overline{\mathcal{V}}^{(k)})$. 

Then the teacher knowledge $\widetilde{Y}_c = [\overline{Y}_c^{(1)},\overline{Y}_c^{(2)},\cdots,\overline{Y}_c^{(K)}]$ can be decoded once the server 
has received sufficient aggregated shares from each group $G_c$, and the central server will extract the real part of the decoded teacher knowledge to transform its mode from $\mathbb{C}^{K \times \Omega \times D}$ to $\mathbb{R}^{K \times \Omega \times D}$.

\subsubsection{Signature Aggregation}
After decoding, the server needs to provide proofs about the 
 execution of the aggregation process to the group leader for verifying the correctness of the teacher knowledge. Based on the weight signatures $\pi^{c}_{\mathcal{W}}= (\pi^c_1, \pi^c_2, \dots, \pi^c_z), z\in G_c$ submitted by the group leader $c$ and the logits signatures $\pi^{c}_{\mathcal{V}} = (\pi'_1, \pi'_2, \dots, \pi'_z)$ submitted by each follower in group $G_c$, the proof $\mathcal{\pi}_c$ is calculated as:

\begin{equation}\label{eq17}
\begin{aligned}
\pi_c=\prod_{z\in G_c}e(\prod_{k=1}^{K}\pi^{(k)'}_z,\pi^c_z)
\end{aligned}
\end{equation}

After completing the computation, the server will send the corresponding information $Res_c = (\widetilde{Y}_c, \pi_c)$ to each leader $c \in [N]$ of group $G_c$.

\subsubsection{Proof Verification}

To obtain the final teacher knowledge, each group leader $c$ need to perform the reverse split process to concatenate the decoded $K$ parts of $\widetilde{Y}_c$.

Case 1: For class-grain type, the leader $c$ performed a summation operation as:

\begin{equation}\label{eq18}
\begin{aligned}
\widetilde{Y}_c=\sum_{k=1}^{K} \overline Y_c^{(k)}, \ \mathbb{R}^{K \times \Omega \times D} \rightarrow \mathbb{R}^{D \times D}
\end{aligned}
\end{equation}

Case 2: For sample-grain type, the leader $c$ performs the row concatenation operation:

\begin{equation}\label{eq19}
\begin{aligned}
\widetilde{Y}_c=\overline Y_c^{(1)} || \overline Y_c^{(2)} ||\cdots || \overline Y_c^{(K)}, \ \mathbb{R}^{K \times \Omega \times D} \rightarrow \mathbb{R}^{O \times D}
\end{aligned}
\end{equation}

Additionally, the blind factor must be removed from $\widetilde{Y}_c$. This is done by applying the matrix $\frac{1}{\textbf{R}_c}$, which consists of the inverses of each element in $\textbf{R}_c$, through the hadamard product as $\hat Y_c=\frac{1}{R_t}\odot \widetilde{Y}_c$.

After deblinding, each leader $c$ verifies the aggregated knowledge $\hat Y_c$ to ensure that there has been no deviation from the SA protocol by the server or followers. Specifically, the server must not tamper with any signatures or decoded knowledge, and followers must not maliciously alter the knowledge splits or shares exchanged during the interaction, while ensuring that the local aggregation is performed according to the agreed-upon weights.

To verify the correctness of the teacher's knowledge, the group leader $c \in [N]$ needs to check if the following equality holds based on the proof $\pi_c$:

\begin{equation}\label{eq20}
\begin{aligned}
e(g,g)^{\sum_{g=1}^M\sum_{l=1}^{D}y_{g,l}+K\sum_{i \in G_c}w_i\Upsilon_i}\overset{?}= \pi_c
\end{aligned}
\end{equation}
where $e(\cdot,\cdot)$ denotes bilinear pairing described in Appendix \ref{app:B}, $y_{g,l}$ represents the $(g,l)$-th element of $\hat{Y}_c$, and $M$ equals to $D$ or $O$ for class and sample grained logits type, respectively. In SVAFD, the use of signature techniques helps reduce the computational overhead for the verifier. This enables participants to independently carry out the verification process, without the need to rely on the third trusted party for assistance.

\section{ANALYSIS}
\label{5}
\subsection{Correctness}
\label{5.1}
The correctness of the SVAFD scheme will be proven around the following questions. 1) Correctness of co-aggregation; 2) Correctness of signature authentication.

\subsubsection{Correctness of Co-aggregation}
\label{5.1.1}
The goal of group leader $c$ is to aggregate the knowledge from its totally $R$ followers in group $G_c$, i.e., the ground truth teacher knowledge is $Y_c^*=\mathcal{W}_c f(\mathcal{\overline{V}}^c)$, and the SVAFD leader $c$ obtains $\hat Y_c=\frac{1}{\mathbf{R}'_c}\odot Dec(\hbar(\widetilde {\mathcal{V}}^c))$, where $\overline{\mathcal{V}}^c=[\overline{\vartheta}_1,\overline{\vartheta}_2,\cdots,\overline{\vartheta}_z]\Leftrightarrow \{\overline{\mathcal{V}}^{(k)}|k \in [K]\}$, and $\widetilde{\mathcal{V}}^c=\{\widetilde{\vartheta}_{\rightarrow z} | z \in G_c\}$.

The correctness of co-aggregation can be proven as follows.
\begin{equation}
\begin{aligned}
& \hat Y_c=\frac{1}{R_c} \odot \operatorname{Dec}\left(\hbar(\widetilde {\mathcal{V}}^c)\right) \\
& \overset{(1)}{=}\frac{1}{R_c} \odot 
\operatorname{Dec}\left(\widetilde{\mathcal{W}}_c f(\operatorname{Enc}(\overline{\mathcal{V}}^c))\right) \\
& \overset{(2)}{=}\frac{1}{R_c} \odot R_c \odot \mathcal{W}_c  f(\operatorname{Dec}(\operatorname{Enc}(\overline{\mathcal{V}}^c)))=Y^*_c
\label{eq_yy}
\end{aligned}
\end{equation}

\textbf{Proof:}
As the data flow diagram illustrated, each follower $z \in G_c$ first split the local knowledge $\vartheta_z$ to get $
\overline{\vartheta}_z=[\overline{\vartheta_{z}}^{(1)},\overline{\vartheta_{z}}^{(2)}, \dots, \overline{\vartheta_{z}}^{(K)}]$, perform LCC encoding operation $Enc(\cdot)$ for $
\widetilde{\vartheta}_{z \rightarrow }=[\widetilde \vartheta_{z\rightarrow 1},\widetilde \vartheta_{z\rightarrow 2},\allowbreak \cdots,\widetilde \vartheta_{z\rightarrow }]$  and exchanges the encoded shares. The collection of all received shares forms the encoded knowledge set $\widetilde{\mathcal{V}}^c$, and the above process can be simplied as $\widetilde{\mathcal{V}}^c=\operatorname{Enc}(\overline{\mathcal{V}}^c)$.  Combining with Eq. \ref{eq5}, we can deduce the first equation in Eq. \ref{eq_yy} holds

According to the Lagrange encoding function in Eq. \ref{lcc}, $u_z(\alpha_x), x \in G_c$ is a polynomial of maximum degree $(K + T - 1)$ with respect to $\alpha$, satisfying that $u_z(\alpha_x) = \widetilde{\vartheta}_{z \rightarrow x}$ and  $u_z(\beta_k) = \overline{\vartheta}_z^{(k)}, k \in [K]$. Defining the set $u(x)=[u_1(x);u_2(x);\cdots;u_z(x)]$.Thus:
\begin{equation}\label{eq22}
\begin{aligned}
u(\Xi) = 
\begin{cases} 
    \widetilde{\vartheta}_{\rightarrow \mathcal{I}(\Xi)}, & \Xi \in \{\alpha_x| x \in G_c \}  \\
    \overline{\mathcal{V}}^{(k)}, &  \Xi \in \{\beta_k| k \in [K] \}
\end{cases}
\end{aligned}
\end{equation}
where $\mathcal{I}(x)$ denotes the subscript of $x$. Since $\hbar(\widetilde {\vartheta}_{\rightarrow z}) = \hbar(u(\alpha_z)) = \widetilde{\mathcal{W}}_c f(u(\alpha_z))$, which poses a degree of $deg(f \times u)=D\times (K+T-1)$. When the aggregation results $\hbar(\widetilde {\mathcal{V}}^c) = \{\hbar(\widetilde{\vartheta}_{\rightarrow z}) \mid z \in G_c \}$ is submitted to the central server, at least $D \times (K + T - 1) + 1$ values are required to reconstruct the polynomial $\hbar(u(\cdot))$ via Lagrange polynomial interpolation. Then $\widetilde{Y}_c$ can be induced by evaluating $\hbar(u(\beta_k)), \ k\in [K]$. The above decoding process can be expressed as $\operatorname{Dec}(\hbar(\widetilde {\mathcal{V}}^c)) = \hbar(\{\overline{\vartheta}^{(k)} \mid k \in [K] \}) = \widetilde{\mathcal{W}}_c f(\{\overline{\vartheta}^{(k)} \mid k \in [K] \}) = R_c \odot\mathcal{W}_c f(\overline{\mathcal{V}}^c)$. Since $\operatorname{Dec}(\operatorname{Enc}(x)) = x$, the second equation of Eq.\ref{eq_yy} is holds, thus the Definition 1 is proved.

\subsubsection{Correctness of Signature Authentication}
\label{5.1.2}
This section provides the verification of the linear weighted aggregation for FD, which is based on the Verifiable Linear Computing process of \cite{liuPrivacyPreservingVerifiableOutsourcing2023}. The formal proof given by the following equation:
\begin{equation}\label{eq23}
\begin{aligned}
\pi_c &=\prod_{i\in G_c}e(\prod_{k=1}^K \pi^{(k)'}_i,\pi^c_i)   =\prod_{i\in G_c}e(g^{\sum_{k=1}^K (V_i^{(k)}+\Upsilon_i)},g^{w_i}) \\
&=\prod_{i\in G_c}e(g,g)^{{(\sum_{k=1}^K\sum_{g=1}^{\Omega}\sum_{l=1}^{D}\ {v_{i,gl}^{(k)}}{w_i}}+K\Upsilon_i{w_i}) } \\
&=e(g,g)^{\sum_{i\in G_c}w_i\sum_{k=1}^K{\sum_{g=1}^{\Omega}\sum_{l=1}^{D}\ {v_{i,gl}^{(k)}}} }e(g,g)^{K\sum_{i\in G_c}w_i\Upsilon_i}\\
&=e(g,g)^{{\sum_{g=1}^{M}\sum_{l=1}^{D}\ {y_{g,l}}} +{K\sum_{i\in G_c} w_i\Upsilon_i} }\\
\end{aligned}
\end{equation}
where $y_{g,l}$ represents the $(g, l)$-th element of the teacher's knowledge $\hat Y_c$, $v_{i,gl}^{(k)}$ denotes the $(g,l)$-th element of $\overline{\vartheta}_i^{(k)}$.

Any clients or the server, who has dishonest behavior during computation and is not detected, needs to make equation $\pi_c\overset{?}{=}e(g,g)^{{\sum_{g=1}^{M}\sum_{l=1}^{D}\ {y_{g,l}}} }e(g,g)^{\sum_{i\in G_c} w_i\Upsilon_i}$ hold. However, for the client-side, it receives the masked 
$\widetilde{\mathcal{W}}_c$ and does not know the decoded value of $\hat{Y}_c$ or the private keys $\Upsilon_i, i \in G_c$ of other participants. Similarly, the server only poses the blinded $\widetilde{Y}_c$. They cannot deceive the leader into accepting tampered teacher knowledge by forging the verification proof $\pi_c$.

\begin{figure*}[htbp]
  \centering
  \includegraphics[width=1\linewidth]{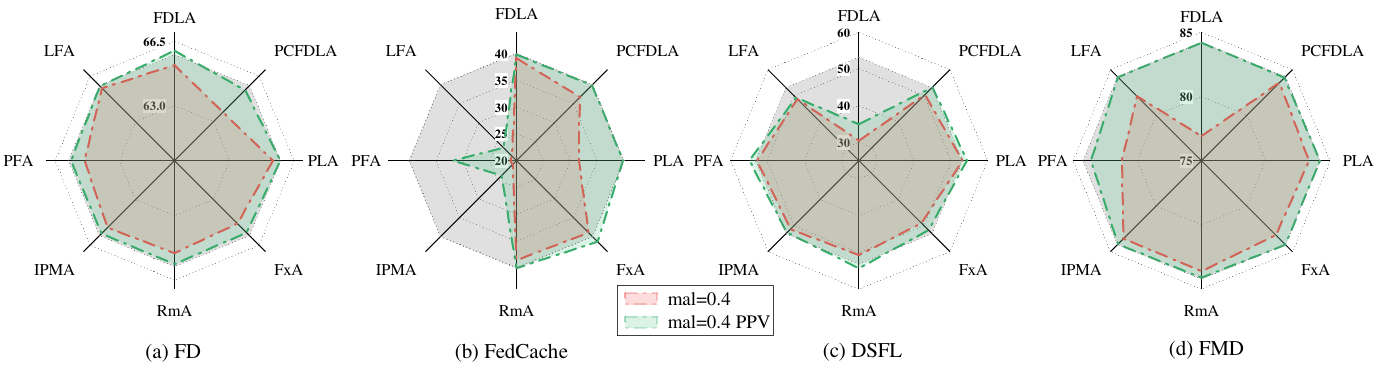}
  \caption{The MAUA of the client model obtained by different aggregation architectures under eights attacks with 40\% malicious clients on SVHN.}
  \label{acc_svhn_40}
\end{figure*}

\begin{figure*}[htbp]
  \centering
  \includegraphics[width=1\linewidth]{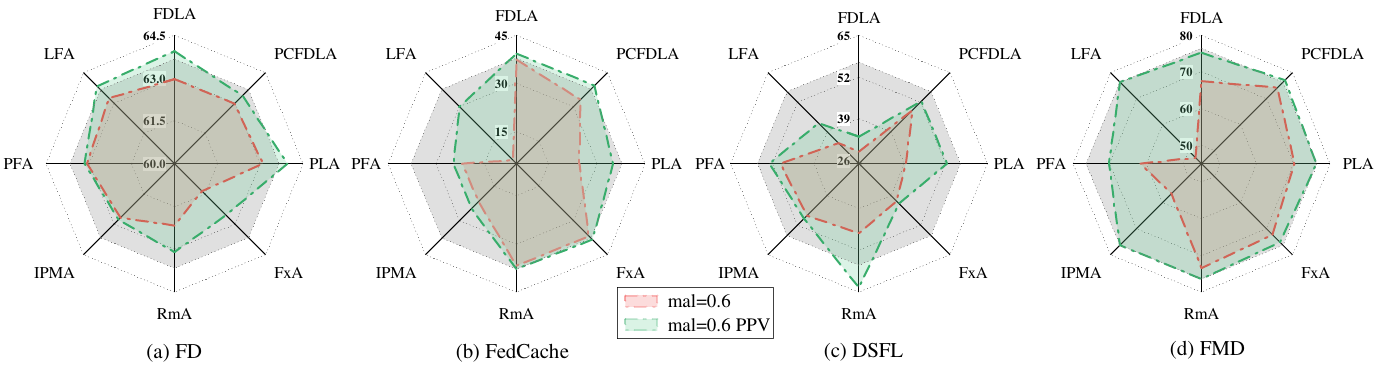}
  \caption{The MAUA of the client model obtained by different aggregation architectures under eights attacks with 60\% malicious clients on SVHN.}
  \label{acc_svhn_60}
\end{figure*}

\section{EXPERIMENTS}
\subsection{Experiment Setup}
\textbf{Implementation.} 

We evaluate the performance of our proposed SVAFD with FedML research library \cite{he2020fedmlresearchlibrarybenchmark} deployed on CUDA version 12.4 for all experiments. We use a random projection matrix \cite{bingham2001random} as the LSH mapping function\cite{10.5555/645925.671516}, use the SS512 curve from the Charm library\cite{akinyele2013charm} as the base for the pairing group, and complete the signature and aggregation operations as described in \cite{liuPrivacyPreservingVerifiableOutsourcing2023},

\begin{table}[t]
  \caption{The configurations of three adopted models. The height and width of the input images are noted as H and W, respectively.}
  \centering
  \begin{tabular}{l c c c}
    \toprule
    \textbf{Model} & \textbf{Notation} & \textbf{Feat. Shape} & \textbf{Params} \\
    \midrule
    ResNet8-small & $A^{C}_{1}$ & $H \times W \times 16$ & 76.2K \\
    ResNet16-medium & $A^{C}_{2}$ & $H \times W \times 16$ & 171.2K \\
    ResNet20-large & $A^{C}_{3}$ & $H \times W \times 16$ & 266.1K \\
  \bottomrule
  \end{tabular}

  \label{table:model_specs}
\end{table}
\textbf{Datasets,Models.} Two typical datasets from different domains in our experiments are utilized, including  SVHN \cite{netzer2011reading} and FashionMNIST \cite{netzer2011reading}. In addition, we consider heterogeneous model architectures {$A_1^C$, $A_2^C$, $A_3^C$} for different clients, and the main configurations of the three adopted models are the same as in \cite{wuFedCacheKnowledgeCachedriven2023a}, as shown in Table \ref{table:model_specs}. Unless otherwise specified, the number of clients in the experiments is 100, using the SGD optimizer with a learning rate of 0.01, batch size of 32, and all client models are selected sequentially from the three types using modulo operations. 

\textbf{Distillation architectures.}
We apply our SVAFD scheme to FD\cite{jeongCommunicationEfficientDeviceMachine2018}, FMD\cite{liFedMDHeterogenousFederated2019a}, and the more challenging novel frameworks Fedcache\cite{wuFedCacheKnowledgeCachedriven2023a} and DSFL\cite{itaharaDistillationBasedSemiSupervisedFederated2023}.

\textbf{Evaluation Metrics.}
The precision of algorithms is measured by Maximum Average User model Accuracy \cite{wuFedCacheKnowledgeCachedriven2023a} (MAUA). Moreover,
we utilized the Attack Success Rate (ASR) as evaluation metrics, while measures the ability of SVAFD to resist attacks. Therefore, a higher MAUA indicates a more effective model, while a lower ASR indicates greater robustness of the model against attacks.

\begin{figure}[!t]
  \centering
  \includegraphics[width=1\linewidth]{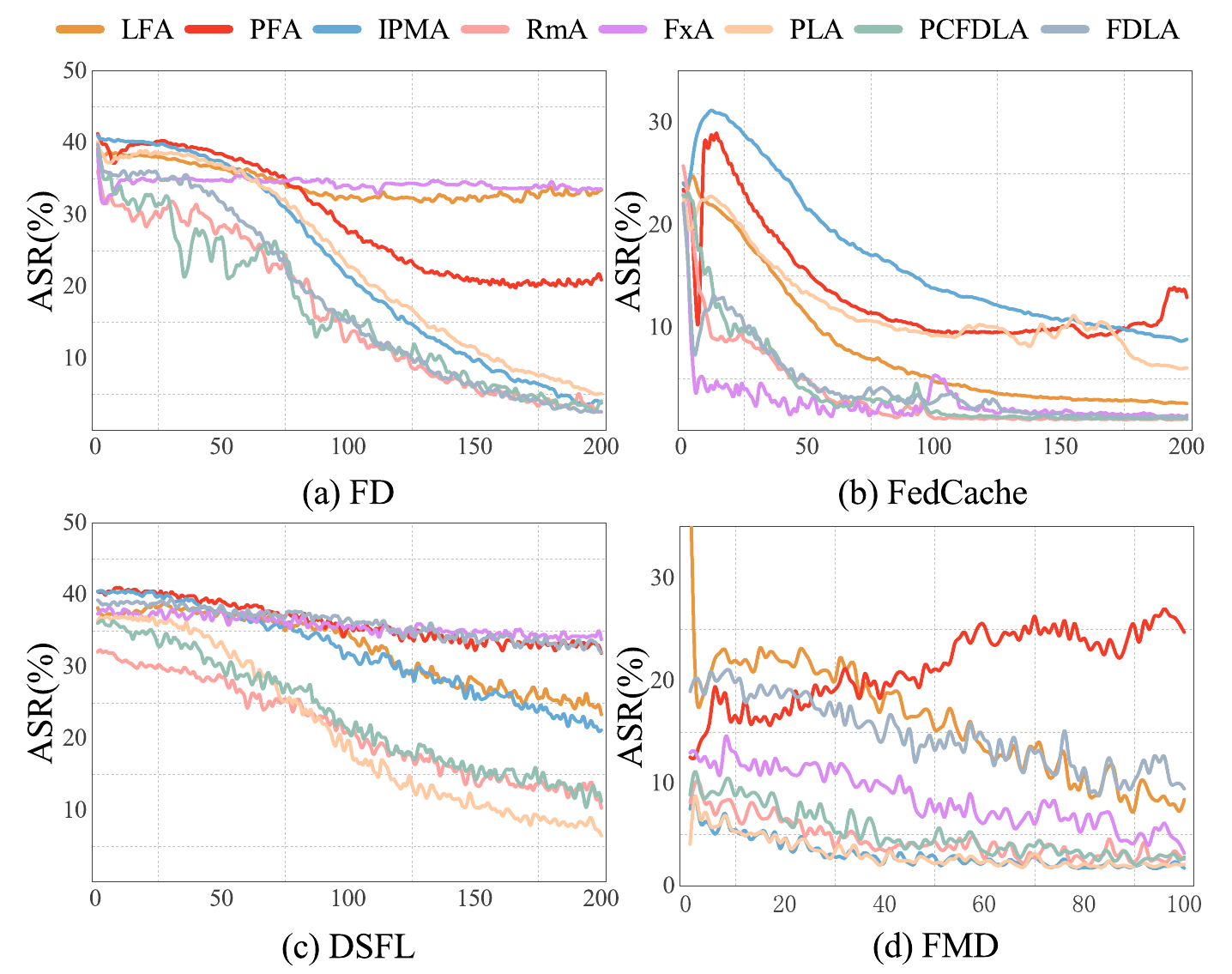}
  \caption{The ASR obtained on SVHN.}
  \label{asr_svhn}
\end{figure}

\subsection{Robustness}
We quantitatively show that SVAFD helps the four FD architectures achieve satisfactory performance while facing eight attack methods targeting data, models, and logits poisioning attacks.
\subsubsection{MAUA under Various Poisoning Attacks}
We evaluate the MAUA of four FD architecture under eight poision attacks over three datasets. Specifically, we consider that SVAFD cooperative with the four benchmarks under (i) the model heterogeneity among clients, and those with residuals of index mod 3 of 0, 1 and 2 are assigned with model architectures $A_1^C$,$A_2^C$ and  $A_3^C$ respectively; (ii) the data heterogeneity on no-iid data distribution, with $\alpha=1$ setting for Dirichlet distribution; (iii) the various attacks, with three different percentages of malicious clients(MP) (40\%, 50\%, 60\%).  We investigate nearly 700 different combinations of approaches, attacks, and tasks. Each combination is trained for a fixed number of 100 epochs for FMD and 200 epochs for the remains three architectures in this section.

Fig. \ref{acc_svhn_40} and Fig. \ref{acc_svhn_60} illustrate the radar charts of the four distillation architectures under the proposed SVAFD when facing various types of poisoning attacks. The red area represents the MUAU of the original distillation scheme, while the green area represents the MUAU performance of the distillation architecture after applying SVAFD. The gray area serves as the reference MUAU without any poisoning attacks. As shown, all four frameworks achieved performance improvements across the eight poisoning schemes, with the average percentage performance improvements for the eight schemes  being 8.54\%, 7.78\%, 21.20\%, 3.09\%, 11.28\%, 12.79\%, 8.40\%, and 29.19\%, respectively.
\begin{figure}[!t]
  \centering
  \includegraphics[width=1\linewidth]{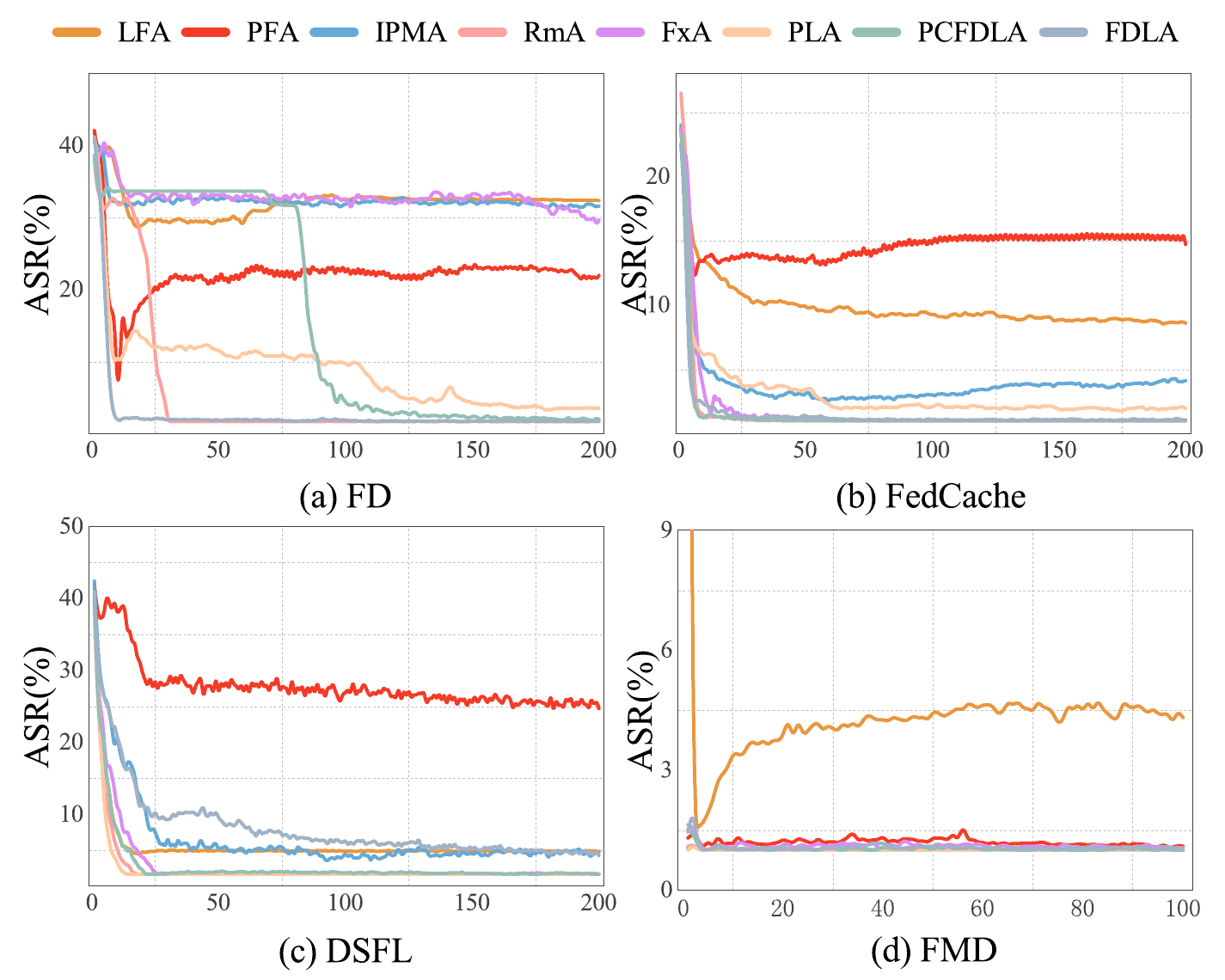}
  \caption{The ASR obtained on FMNIST.}
  \label{asr_fmnist}
\end{figure}

\begin{figure*}[htbp]
    \centering
    \begin{subfigure}[b]{1\linewidth} 
        \centering
        \includegraphics[width=\linewidth]{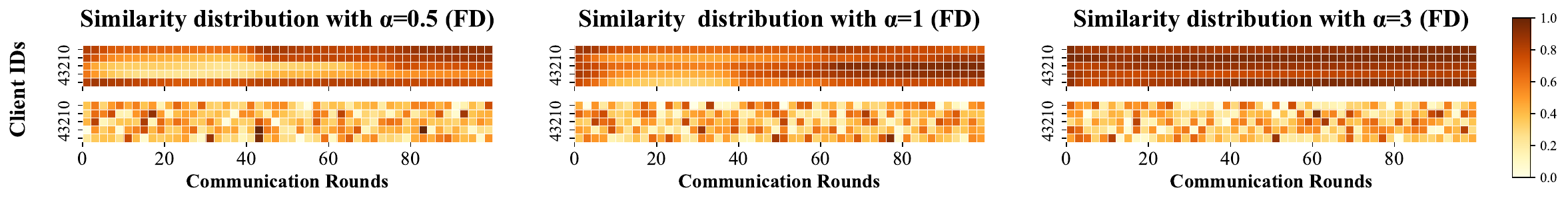}
        \caption{The similarity of knowledge distribution and label distribution (FD).}
        \label{subfig:hot_FD}
    \end{subfigure}
    \vspace{1em}
    \begin{subfigure}[b]{1\linewidth} 
        \centering
        \includegraphics[width=\linewidth]{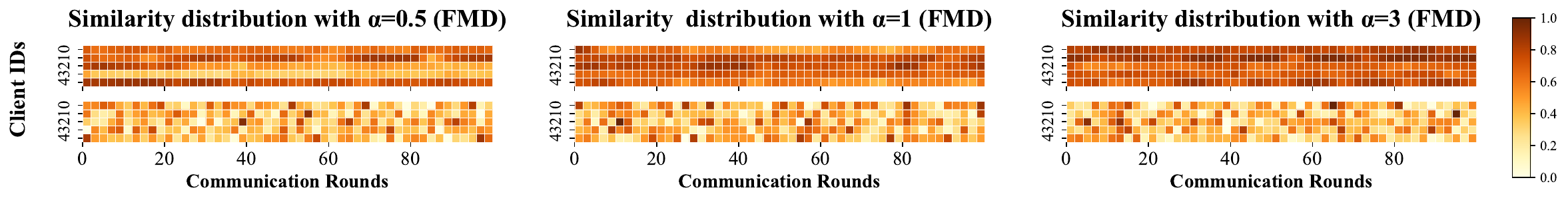}
        \caption{The similarity of knowledge distribution and label distribution (FMD).}
        \label{subfig:hot_FMD}
    \end{subfigure}
    \caption{The similarity of knowledge distribution and label distribution in different cases.}
    \label{fig:hot_similarity}
\end{figure*}

The divergencey in the areas of the green and red regions visually demonstrates the robustness of the SVAFD scheme against various poisoning attacks. In the combinations where MUAU differences are significant, such as RMA in Fig.\ref{acc_svhn_60}(c), and LFA in Fig.\ref{acc_svhn_60}(d), SVAFD helps the corresponding architecture achieve the 36.46\% and 61.11\% accuracy improvement. This occurs when the number of aggregated clients is only 40\% of that in the no-attack scenario, showing performance that is comparable to, or even superior to the gray area. This is because SVAFD continuously updates the clients' intimacy list during training with the quality-aware scheme, allowing each client to carefully select efficient teacher knowledge. SVAFD effectively avoids the inconsistency of knowledge and the impact of poisoning attacks on model convergence, while reducing network resource usage by aggregating only a portion of the clients rather than all candidate clients.

\subsubsection{ASR under Various Poisoning Attacks}
We evaluate the ASR of four architecture with 60\% malicious clients over SVHN and FMNIST datasets. As shown in Fig.\ref{asr_svhn}, we observe that the success rate of all attack methods decreases continuously as training progresses and gradually stabilizes. By the end of training, the success rate of most attack methods has dropped below 20\%. This is because, as the local model accuracy improves, the client model's ability to fit the local data feature distribution strengthens. As a result, the difference between clients with consistent knowledge distributions and malicious clients becomes more pronounced, making it easier for the quality-aware scheme to filter out candidates.

Fig.\ref{asr_fmnist} also includes the performance of SVAFD on the FMNIST dataset. The ASR converges to below 10\% in DSFL and FMD within the first 20 rounds, showing superior performance compared to the SVHN dataset. This might be because the FMNIST task is relatively simpler, and the rapid improvement in model accuracy facilitates the quality-aware scheme's efficient filtering of malicious clients. LFA and PFA exhibit higher ASR compared to other attack methods. Nevertheless, SVAFD still reduces more than 70\% of these two malicious attacks, effectively enhancing the robustness of the FD system.

\begin{figure*}[htbp]
    \centering
    \begin{minipage}{0.30\textwidth}
        \centering
        \includegraphics[width=\textwidth]{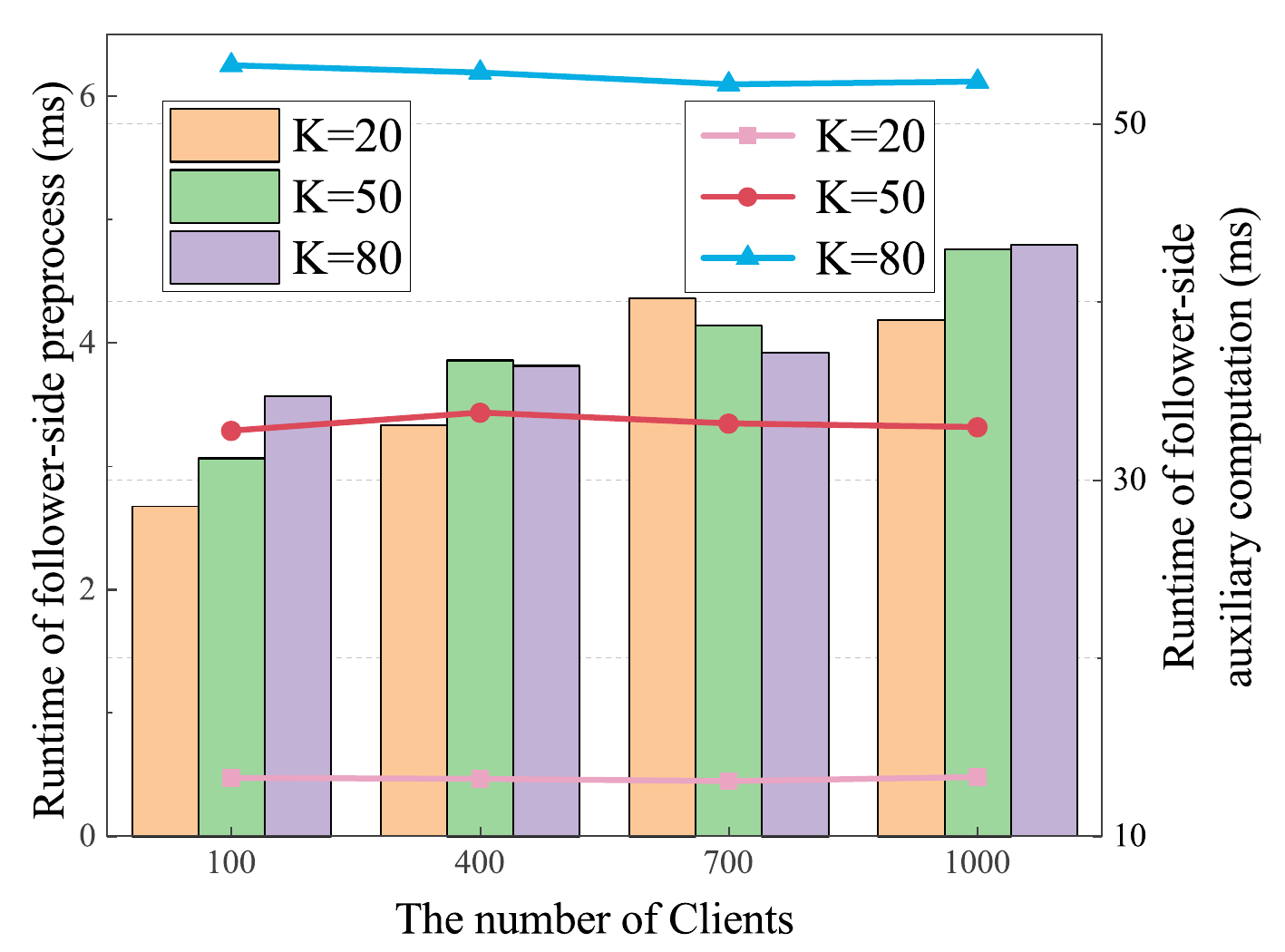}
        \caption{Run time of preprocess and auxiliary for followers.}
        \label{s1}
    \end{minipage}%
    \hfill
    \begin{minipage}{0.30\textwidth}
        \centering
        \includegraphics[width=\textwidth]{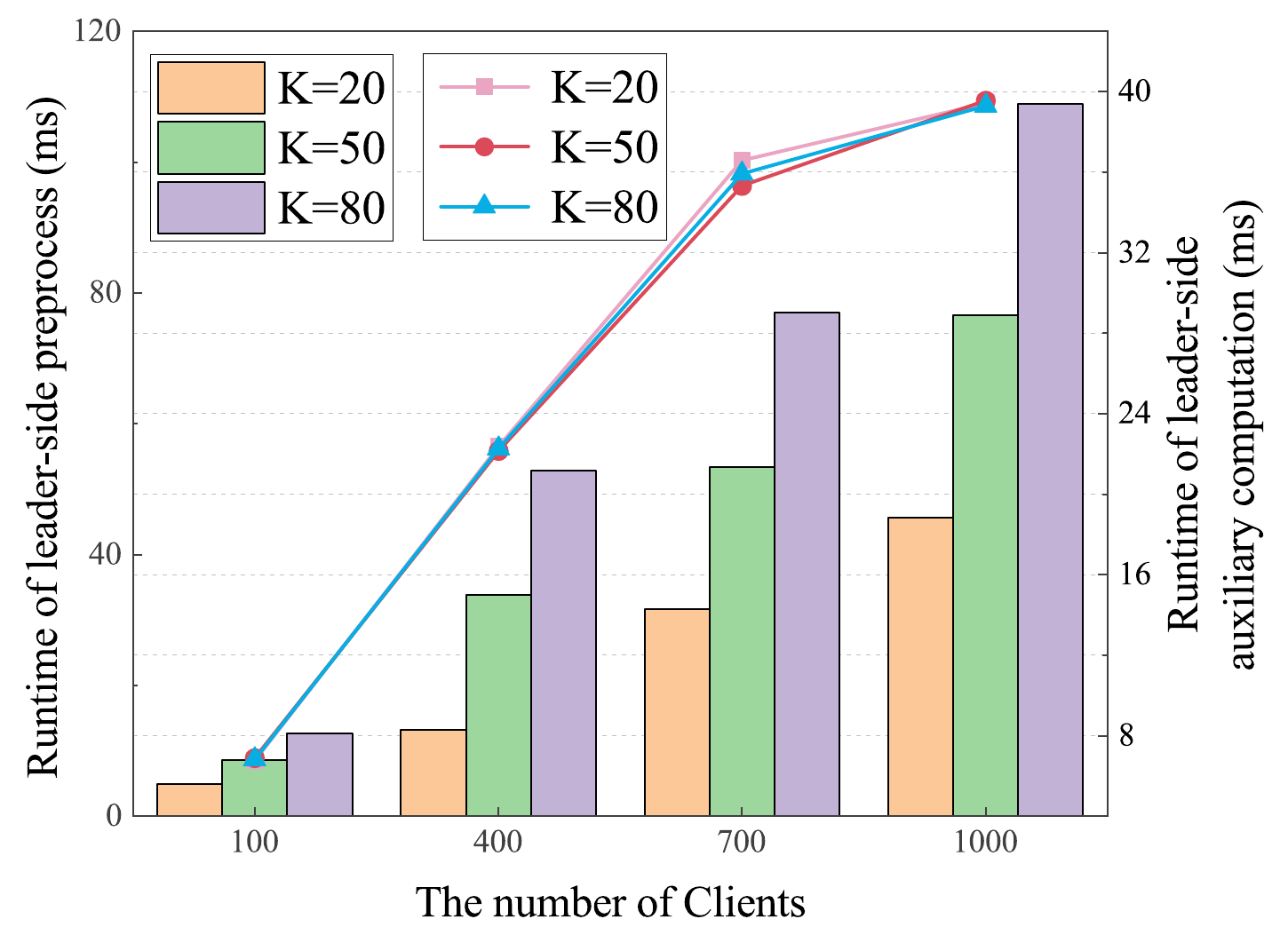}
        \caption{Run time of preprocess and auxiliary for leaders.}
        \label{s2}
    \end{minipage}%
    \hfill
    \begin{minipage}{0.30\textwidth}
        \centering
        \includegraphics[width=\textwidth]{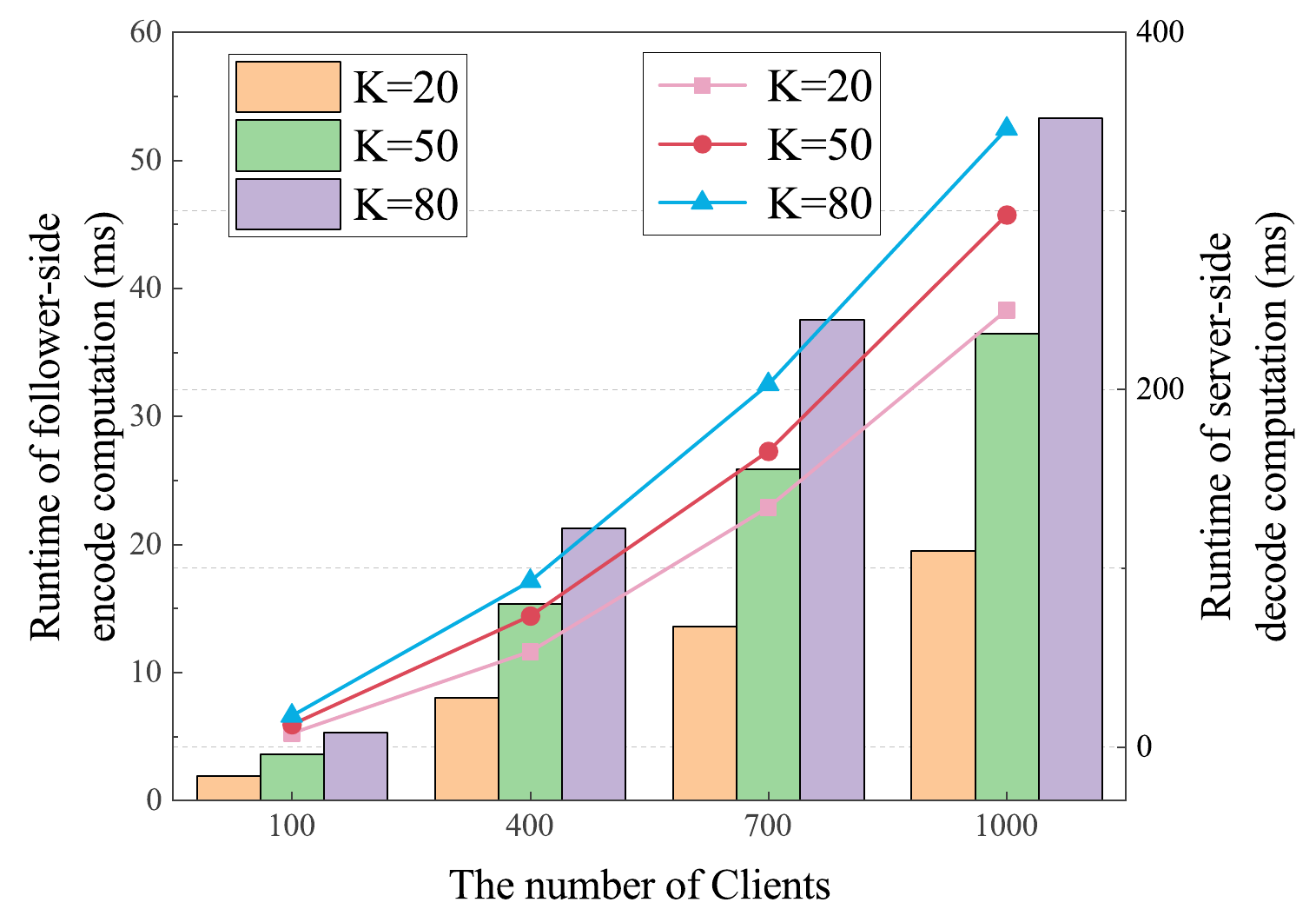}
        \caption{Run time of encode and decode in class-grained.}
        \label{s3}
    \end{minipage} \\
    
    \begin{minipage}{0.30\textwidth}
        \centering
        \includegraphics[width=\textwidth]{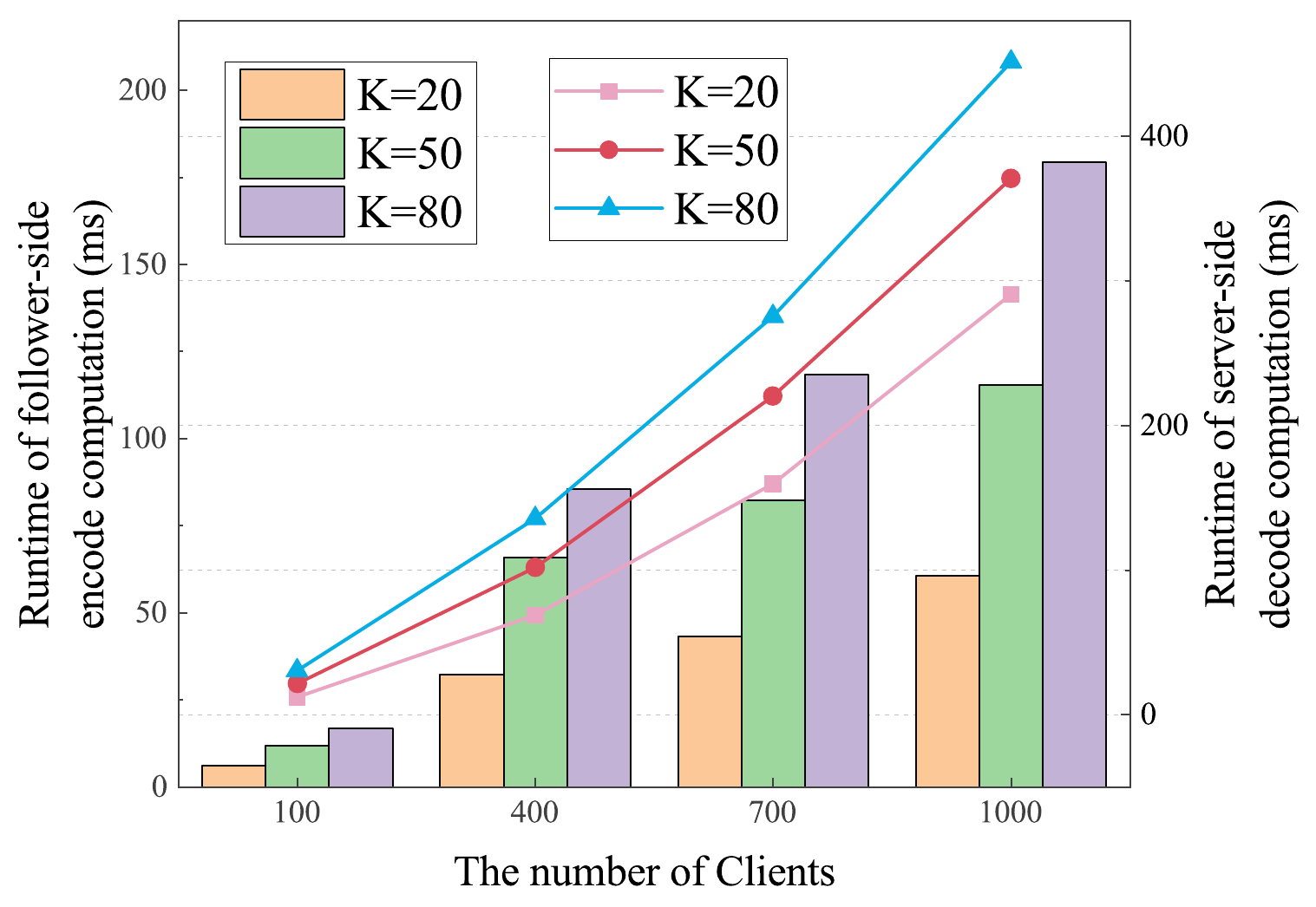}
        \caption{Run time of encode and decode in sample-grained.}
        \label{s4}
    \end{minipage}%
    \hfill
    \begin{minipage}{0.30\textwidth}
        \centering
        \includegraphics[width=0.97\textwidth]{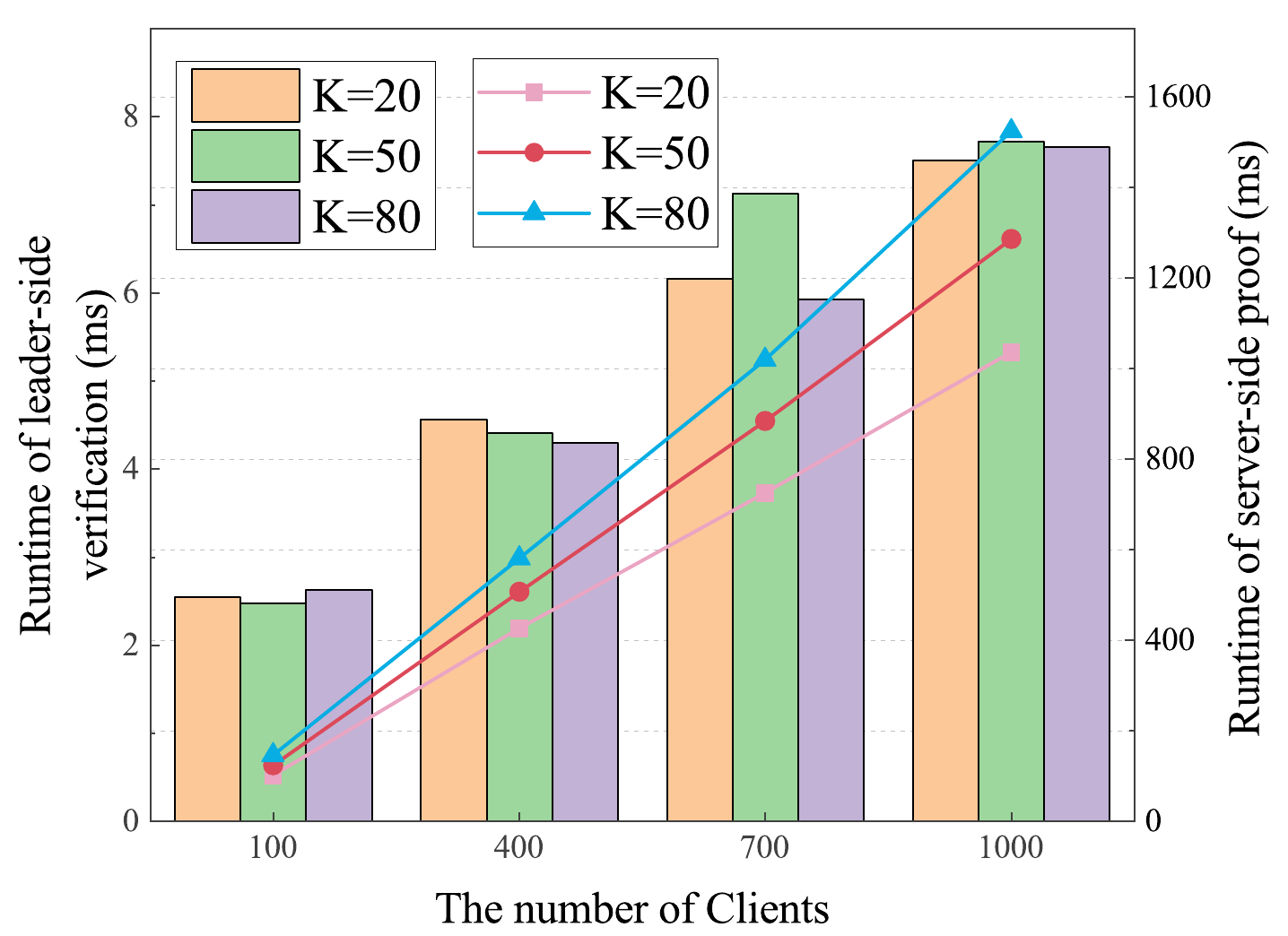}
        \caption{Run time of verification and proof generation.}
        \label{s5}
    \end{minipage}%
    \hfill
    \begin{minipage}{0.3\textwidth}
        \centering
        \includegraphics[width=0.88\textwidth]{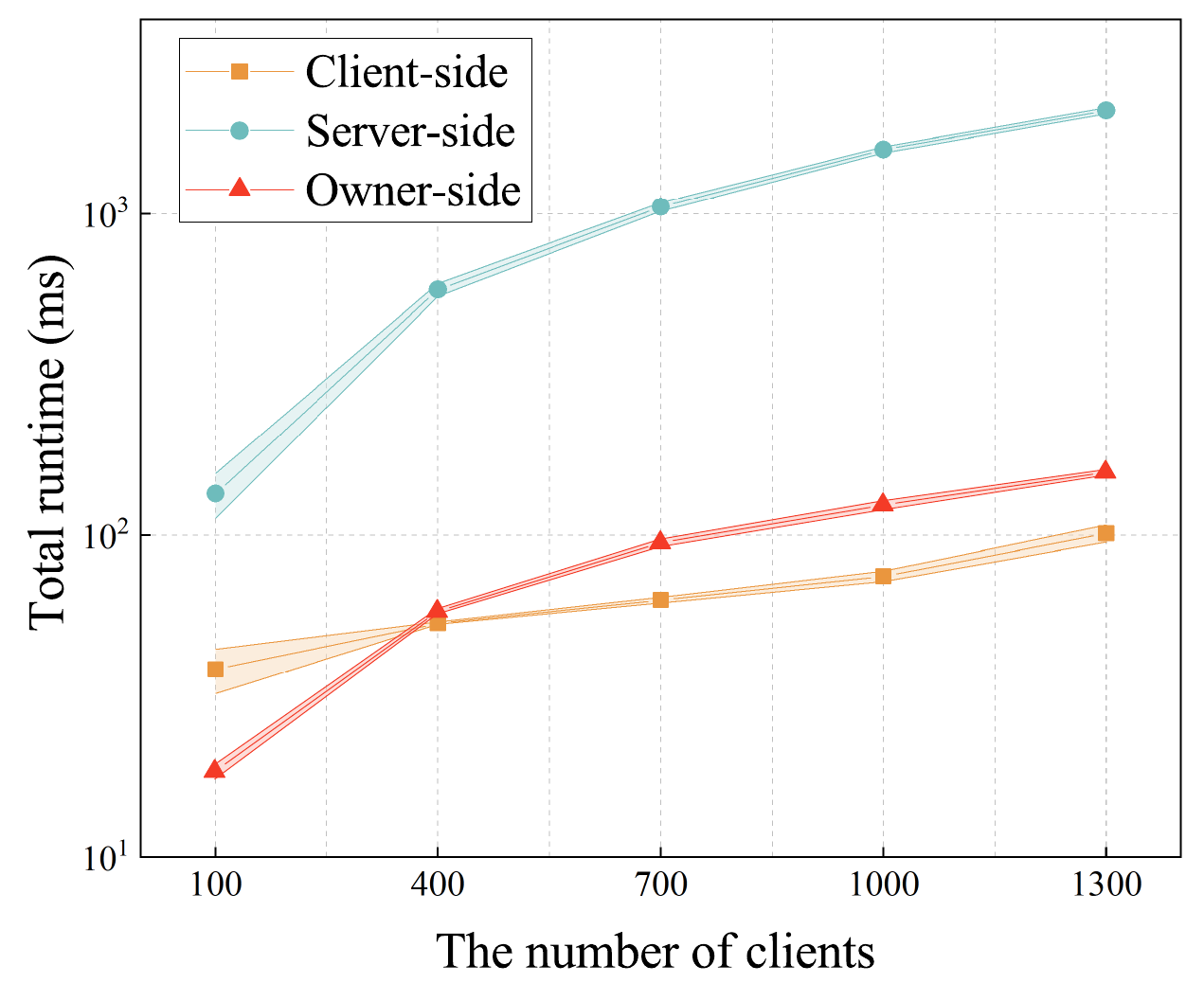}
        \caption{Total run time of participants with T=50.}
        \label{s6}
    \end{minipage}
\end{figure*}

\subsubsection{Data Distribution Security}
To explore the data distribution security of SVAFD towards the DDI, we set the Dirichlet distribution hyper-parameter to control the degree of local data distribution, and compare the performance of SVAFD in the class-grained(FD) and sample-grained(FMD) scenarios. We take five clients as an example, and the heatmap describes the change in the similarity relationship $\eta_d$ between the mean logits distribution of C/S interactions and the local data distribution over the first $100$ rounds of training. 

In Fig.\ref{fig:hot_similarity}, the upper part of each sub-heatmap shows the change in $\eta_d$ without any defensive measures. As shown, with the training progresses, the color gradually deepens, and the approximate value of $\eta_d$ continues to rise, stabilizing at a high correlation. This is because the model's training goal is to minimize the difference between the predicted distribution and the true label distribution through optimization (e.g., cross-entropy loss). After sufficient training, the model's logits output statistically aligns with the label distribution of the training data. When the SVAFD scheme is applied, the lower part of the heatmap shows a more random distribution. This occurs because the teacher knowledge obtained by the server about the client is based on the aggregation of knowledge from $R$ clients which is randomly masked. In this case, the server cannot capture the local data distribution information based on the mean logits distribution from the interactions.

\subsection{Efficiency}
Our evaluations about the efficiency are centered around the relative error of the encoding, as well as the latency at different stages.

\begin{table}[t]
    \caption{The relative error RE (log10) under different parameter combinations with $\sigma=10^3$, $\theta=6$, batchsize=32, $\beta=1.15$.}
    \centering
    \renewcommand{\arraystretch}{1}  
    \scalebox{0.65}{
    \begin{tabular}{c|ccc|ccc|ccc}
        \toprule  
        \multirow{2}{*}{N} & \multicolumn{3}{c|}{K=10} & \multicolumn{3}{c|}{K=20} & \multicolumn{3}{c}{K=30}  \\
         &T=10  & T=20 & T=30 &T=10  & T=20 & T=30&T=10  & T=20 & T=30 \\
        \midrule 
        50 & -10.75 & -11.08 & -10.98 & -11.03 & -10.71 & -10.57 & -10.80 & -10.80 & N/A \\
        75 & -9.24 & -9.30 & -9.65 & -9.71 & -9.74 & -9.28 & -9.70 & -9.029 & -9.43 \\ 
        100 & -8.03 & -7.42 & -7.96 & -7.87 & -8.01 & -7.49 & -7.78 & -7.38 & -8.08 \\ 
        \bottomrule
    \end{tabular}
    }
    \label{error}
\end{table}

\subsubsection{Relative Error Introduced by MM-LCC}
In this section, we use the relative error $RE=\frac{\| \hat Y-Y^* \|}{\| Y^* \|}.$ to study the error loss after the LCC encoding, aggregation, and decoding processes. Table.\ref{error} illustrates the log values of the relative error under different combinations of the number of clients, splits, and privacy guarantees. All values are the averages from 5 repeated runs. As shown, the relative error does not vary significantly with changes in $K$ and $T$. As the number of clients $N$ increases, the cumulative error caused by encoding operations on all clients grows. However, under the unified condition of $N=100$ in our former experiment, the relative error remains at the order of $10^{-7}$, which indicates that the SVAFD scheme, after receiving a sufficient number of encodings, results in almost negligible decoding errors.
The results fully demonstrates the high precision of our SVAFD.

\subsubsection{System-level Overhead}
In this segment, we study the system-level overhead of SVAFD, including the individual latency of participants in the three stages.

In the first stage, both the leader and followers need to perform data initialization and generate auxiliary signature information. On the follower-side, the initialization overhead is primarily due to hash mapping, knowledge splitting, and knowledge perturbation, while on the leader-side, the primary tasks involve Lagrange matrix computation and weight masking operations. Additionally, both parties need to perform the necessary data quantization and signing.

As shown in Fig.\ref{s1}, the initialization latency on the follower-side is mainly influenced by the number of clients, while the auxiliary computation overhead increases as the value of $K$ grows. The latter's overhead can be up to nearly 10 times higher than the former ($N$=1000, $K$=80). This is because the signing operation involves mapping the quantized logits to an elliptic curve for large integer operations, which inevitably introduces computational overhead. Since $K$ determines the splits of data, as $K$ increases, each client needs to compute more signatures. Even so, under the setting of $K$=80 with $N$=1000 clients, the latency for generating auxiliary information on the client side remains in the millisecond range.

Fig.\ref{s2} shows the overhead for initialization and auxiliary computation on the leader-side. The initialization overhead of the leader is relatively high and increases with the growth of the client number $N$ and the value of $K$. Since the auxiliary computation overhead mainly comes from the quantization and signing, it is primarily influenced by the number of clients and does not fluctuate significantly with changes of $K$. Additionally, we observe that the leader overhead increases nearly linearly with the number of clients.

Fig.\ref{s3} and Fig.\ref{s4} show the encoding overhead on the follower-side and decoding overhead on the server-side under both class-grained and sample-grained types. The overall latency overhead of the sample-grained framework is higher than that of the class-grained framework. Specifically, the decoding latency and encoding overhead in the sample-grained type are on average nearly $3.2$ times and $1.3$ times greater, respectively, than those in the class-grained setup. This is because the size of each split in the sample-grained framework is $\Omega \times D$,relating to the sample numbers, whereas in the class-grained framework, it is  $D \times D$ and $\Omega \geq D$. It's observed that the experimental overhead increases linearly with the values of $K$ and $N$. This is because LCC is inherently a lightweight linear encoding scheme\cite{yuLagrangeCodedComputing2019}, which holds promise for handling large-scale data transmission, providing fault tolerance and optimizing network bandwidth utilization.

Fig.\ref{s5} presents the time taken by the server to generate the aggregation proof and the time for leaders' authentication and deblinding. It can be observed that, even under the most complex scenario with $K=80$ and $N=1000$, the client authentication process is still completed within $10ms$. In contrast, the server’s proof generation time is nearly $2000$ times longer than the authentication time. This demonstrates that relying solely on resource-constrained local clients to perform the signature aggregation required for privacy protection is unrealistic. However, SVAFD addresses this by offloading the heavy signature aggregation task to the computationally resource-rich server, and with the help of the lightweight linear encoding scheme, it achieves efficient computing, thus effectively ensuring user privacy during collaborative training.

SVAFD takes into account the actual demands and resource capabilities of all participants, ensuring that each party can bear the computational overhead. Fig.\ref{s6} provide detailed overview of the overhead. Under both the sample and class types, there is a significant difference in the overhead for encoding and decoding processes. Therefore, additional curves are used to distinguish the performance of client-side and server-side computations. The error ranges from five repeated runs are shown using variance bands. The server-side computational overhead is, on average, $10$ times and $12$ times greater than that of follower-side and leader-side, respectively. Despite this, the server can still complete the computation within $2$ seconds with 1300 clients, while the client-side requires only millisecond-level runtime. This indicates that SVAFD is a lightweight secure aggregation protocol that ensures efficient execution of the training process by keeping the computational overhead within acceptable limits for all parties.

\section{Conclusion}
We propose SVAFD, a verifiable co-aggregation framework for Federated Distillation (FD). SVAFD addresses critical security challenges in FD by decoupling the aggregation process into knowledge selection, aggregation, and verification, ensuring privacy protection and knowledge integrity even in the presence of a malicious server. By leveraging Lagrange Coding Computation (LCC), SVAFD enables efficient and secure co-aggregation in resource-constrained environments such as mobile edge networks. To our knowledge, SVAFD is the first framework to provide secure aggregation specifically for FD, overcoming the limitations of traditional secure aggregation methods that are ill-suited for FD’s heterogeneous model setups. Extensive experiments demonstrate that SVAFD is highly robust against various poisoning and inference attacks, while also significantly improving model accuracy.

\appendix
\section{Appendix A}

\subsection{Details of Two Types of Distillation}
\label{app:A}
\subsubsection{Class-grained Logits Interaction-based Architecture (CLIA)}
For CLIA, the output of each sample $x^c_i \in S^c$ for client $c \in [N]$ needs to be close to the teacher knowledge with the same label\cite{jeongCommunicationEfficientDeviceMachine2018}, that is:

\begin{equation}\label{eq24}
\begin{aligned}
\underset{\Phi^C}{\arg \min } & \sum_{\left(x_i^c, y_i^c\right) \in \mathcal{S}^c}\left[L_{C E}\left(\varkappa_0\left(\ell^c\left(x_i^c\right)\right), y_i^c\right)\right. \\
& \quad + \lambda_c \cdot L_{KD}\left(\varkappa_0\left(\ell^c\left(x_i^c\right)\right), \varkappa_0\left( \vartheta^{g}(y_i^c)\right)\right)]
\end{aligned}
\end{equation}
where $\varkappa_0(\cdot)$ is the softmax mapping function, and $L_{CE}(\cdot)$ represents the cross-entropy loss. $Y^{c}(y_i^c)$ refers to the teacher knowledge from the server with the label $y_i^c$. $\vartheta_c \in \mathbb{R}^{D \times D}$ represents the local average logits of all samples with the same label for client $c$, computed as:

\begin{equation}\label{eq25}
\begin{aligned}
\vartheta_c(y_d)=Avg[\underset{\left(X_i^c, y_i^c\right) \in \mathcal{S}^c \wedge y_d=y_i^c}{\sum} \ell^c\left(X_i^c\right)] 
\end{aligned}
\end{equation}

CLIA supports model heterogeneity through lightweight communication. After each round of local training, the client uploads the average class-grained knowledge $\vartheta_c$ of its local data. This is aggregated by the server to obtain teacher knowledge for the next round of local training.

\subsubsection{Sample-grained Logits Interaction-based Architecture (SLIA)}

For SLIA, it typically requires the introduction of a public dataset or an unlabeled dataset\cite{liFedMDHeterogenousFederated2019a,itaharaDistillationBasedSemiSupervisedFederated2023}. For client $c$, the logits learns from the average logits of all clients on a given sample $(X^o_i, y^o_i)$ from the public dataset $S_o$, i.e.:

\begin{equation}\label{eq26}
\begin{aligned}
\underset{W^c}{\arg \min } & \sum_{\left(X_i^o, y_i^o\right) \in \mathcal{S}^o} L_{C E}\left(\varkappa_0\left(\ell^c\left(X_i^o\right)\right)\right. , \\
& \left.\varkappa_0\left(\vartheta^g(i)\right)\right)
\end{aligned}
\end{equation}

where, $\vartheta^{g}(i)$ represents the teacher knowledge from the public accessible dataset with index $i$. At this point, the teacher knowledge $v^g \in \mathbb{R}^{O \times D}$, where $O$ is the total number of samples selected from the public dataset. The client's local knowledge $\vartheta_c \in \mathbb{R}^{M \times D}$ is expressed as:

\begin{equation}\label{eq27}
\begin{aligned}
\vartheta_c=\underset{\left(X_i^o, y_i^o\right) \in \mathcal{S}^o}{E} \frac{\ell^c\left(X_i^c\right)}{U}
\end{aligned}
\end{equation}
where $U$ is a hyperparameter for distribution control of the aggregated logits. 

\subsection{ Bilinear Pairings}
\label{app:B}
The bilinear pairing group is define as $(p, \mathbb{G}, \mathbb{G}_T, e)$, where $p$ is a large prime number, $\mathbb{G}$ and $\mathbb{G}_T$ are prime-order cyclic multiplicative groups with order $p$, and $e$ is a bilinear map function as $e: \mathbb{G} \times \mathbb{G} \to \mathbb{G}_T$. The bilinear map satisfies the following three properties:

\begin{itemize}
    \item \textbf{Bilinearity}: For all $a, b \in \mathbb{Z}_p$ and $g \in \mathbb{G}$, $e(g^a, g^b) = e(g, g)^{ab}$.
    \item \textbf{Non-degeneracy}: There exists an element $g \in \mathbb{G}$ such that $e(g, g) \neq 1$.
    \item  \textbf{Computability}: For all $g \in \mathbb{G}$, there is an efficient algorithm to compute $e(g, g)$.
\end{itemize}

\subsection{Data Precision Conversion }
\label{app:C}
We take the precision conversion function as $w_z \leftarrow\textbf{Conv}(w_z,q)=\lfloor w_z * 10^{q}\rfloor$, and $V_z^{(k)} \leftarrow\textbf{Conv}(V_z^{(k)},q)=\lfloor V_z^{(k) }* 10^{q}. \rfloor$. This approximation process inevitably results in some information loss and a reduction in precision, however, it does not impact the accuracy of the verification.

\section{Simple References}



\end{document}